\pdfoutput=1

\documentclass[twocolumn,amssymb,floatfix,deluxe,prd,showkeys,nofootinbib]{revtex4-1}

\usepackage{amsmath}
\usepackage{graphicx}
\usepackage{dcolumn}
\usepackage{bm}
\usepackage{epsfig}

\begin{document}

\title{The calorimetric spectrum of the electron-capture decay of $^{163}$Ho.\\
The spectral endpoint region}

%

\author{A. De R\'ujula}
\affiliation{IFT(UAM), Madrid, Spain; CERN,
1211 Geneva 23, Switzerland}%
\author{M. Lusignoli}
\affiliation{Sapienza, Universit\`a di Roma, and INFN, Sezione di Roma
Piazza A. Moro 2, I-00185 Roma, Italy}

\date{\today}

\begin{abstract}
The electron-neutrino mass (or masses and mixing angles) may be directly measurable
in weak electron-capture decays. The favoured experimental technique is ``calorimetric''.
The optimal nuclide is $^{163}$Ho, and
several experiments (ECHo, HOLMES and NuMECS) are currently studying its decay.
The most relevant range of the calorimetric-energy spectrum extends
for the last few hundred eV below its endpoint. It has not yet been 
well measured. We explore the theory, mainly in the cited range, of electron capture in 
$^{163}$Ho decay. A so far neglected process turns out to be most relevant:
electron-capture accompanied by the shake-off of a second electron. Our two main conclusions
are very encouraging: the counting rate close to the endpoint may be more than
an order of magnitude larger than previously
expected; the ``pile-up" problem may be significantly reduced.

\end{abstract}

\maketitle

\section{Introduction} 

Fourscore and three years after Fermi computed
 how a nonzero neutrino mass would affect the endpoint of the electron
spectrum in a $\beta$-decay process \cite{Fermi}, the laboratory quest
for a non-zero result in this kind of measurement is very much alive \cite{Weinheimer}.

Weak electron capture (EC) has a sensitivity to the neutrino mass entirely analogous 
to the one of $\beta$-decay. EC is
the $e\,p\!\to\! \nu\,n$ weak-interaction process whereby an atomic electron interacts with a 
nucleus of charge $Z$ to produce a neutrino, leaving behind a nucleus of charge $Z-1$
and a hole in the orbital of the daughter atom from which the electron was captured.
The optimal nuclide in this respect is $^{163}$Ho. The favoured experimental technique
is ``calorimetric'' \cite{ADR,ADRML,Nucciotti}.
Various experiments --ECHo \cite{ECHo}, HOLMES \cite{HOLMES} and 
 NuMECS \cite{NUMECS}-- are currently making progress in this direction.

Once upon a time it was argued that the calorimetric energy spectrum of $^{163}$Ho decay
 ought to be very well approximated by a simple theoretical expression: the sum
 over the Breit-Wigner-shaped contributions of the single holes left by electrons
  weakly captured by the nucleus \cite{ADRML}. In the
 extremely good approximation in which nuclear-size effects are neglected, 
 the corresponding orbitals are the ones whose wave function at the origin is non-vanishing.
 In $^{163}$Ho EC decay to $^{163}$Dy the 
 reaction's Q-value is smaller than the 
 L (n=2) binding energies so that  
  the potentially capturing orbitals (or resulting holes) are
  H = M1, M2, N1, N2, O1, O2 and P1, after
  which the 67th element (Ho) runs out of electrons.
  
Robertson has pointed out
that two-hole contributions should not be negligible \cite{Robertson1,Robertson2}. 
In an EC event, the wave functions of the spectator electrons in the mother and
daughter atoms are not identical, the small mismatch leading to an ``instantaneous''
creation of secondary holes, H'. An electron having been expelled from the
H' orbital may
be ``shaken up'' to an unoccupied atomic level or
``shaken off'' into the continuum.
The ensuing contribution to the energy distribution would 
result in a peak
for shake-up and a broad feature for shake-off.

The probabilities $P\rm (H,H')$ for the production of two holes
are much smaller than single-hole ones, $P\rm (H)$. Yet, when the energy deficit
of the two-hole state
$E({\rm H,H')}\sim E({\rm H)}+E({\rm H')}$
is not very close to that of a prominent --but narrow--
single-hole peak, there is an observable feature in the
spectrum, even if $P{\rm (H,H')}\ll P\rm (H)$.

Robertson argues that {\it ``If the presence of the
curvature in the spectrum} [due to two-hole processes] 
{\it near the endpoint were not known to an
analyst, fitting to the standard spectral shape would produce erroneous
results for Q and $m_\nu$''}. The two-hole processes, if significant,
will be observed much before the analyst attempts to measure the cited parameters. Moreover,
given the recently measured Q-value, 
$Q\equiv M({\rm ^{163}Ho})-M({\rm ^{163}Dy})=2833\,\rm (30_{stat})\,(15_{sys})$  eV  \cite{Qval},
the dangerous possibility that $E({\rm H,H')}\sim Q$ for any given pair of holes is excluded.

The bad news is that the cited value of $Q$ is larger than the previously `recommended'' 
figure, $Q=2.555\pm 0.016$ keV \cite{Audi}, which goes in the direction of making
a potential measurement of $m_\nu$ more difficult. The good news, as we shall argue,
is that the contribution of two-hole states close to the spectral endpoint may compensate
for the bad news, possibly in an overwhelming way.

We shall be specifically interested in calorimetric energies, $E_c$, in a domain hardly explored
so far, extending from the M1 peak at  $E_c\sim 2050$ eV to the endpoint at $E_c=Q-m_\nu$.
Though QED and the weak-interaction theory are well established to impressive levels of
precision, dealing with atoms containing up to 67 electrons is not entirely straightforward.
Thus, we shall need observational input to guide our course.

\section{Previous results}

\subsection{Data vs theory}

\begin{figure}[htbp]
\begin{center}
\includegraphics[width=0.50\textwidth]{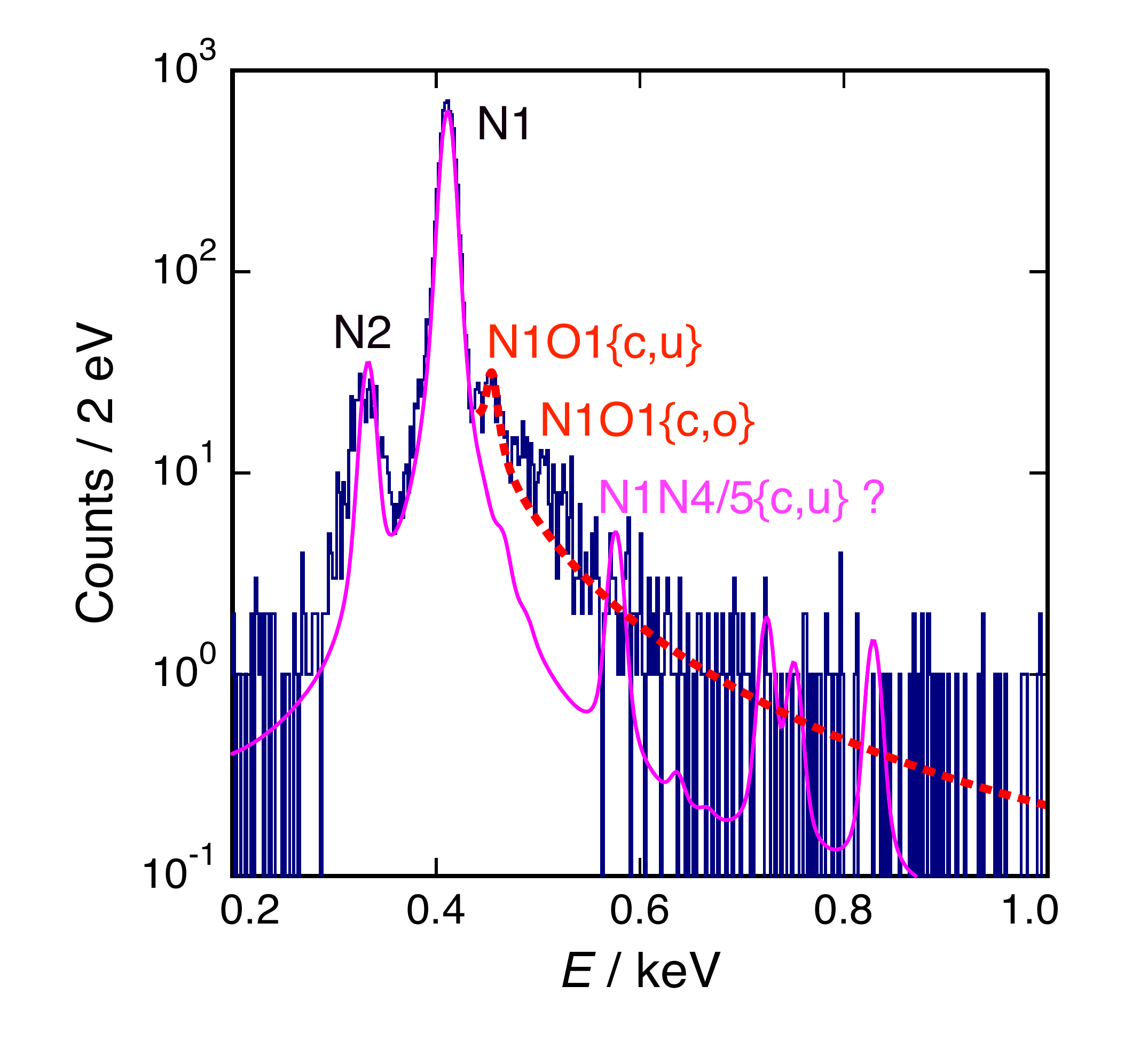}
\end{center}
\caption{Blue: calorimetric spectrum measured by ECHo \cite{ECHo2}.
Magenta: prediction of Faessler et al.\,\cite{Faessler2}.
Red, dotted: added effect of the contributions N1O1\{c,u\}
(arbitrarily multiplied by 2.5) and N1O1\{c,o\} \cite{us}.
The notation \{c,u(o)\} indicates that one electron is captured,
one shaken up (off).
}
\label{fig:LG}
\end{figure}

\begin{figure}[htbp]
\begin{center}
\includegraphics[width=0.48\textwidth]{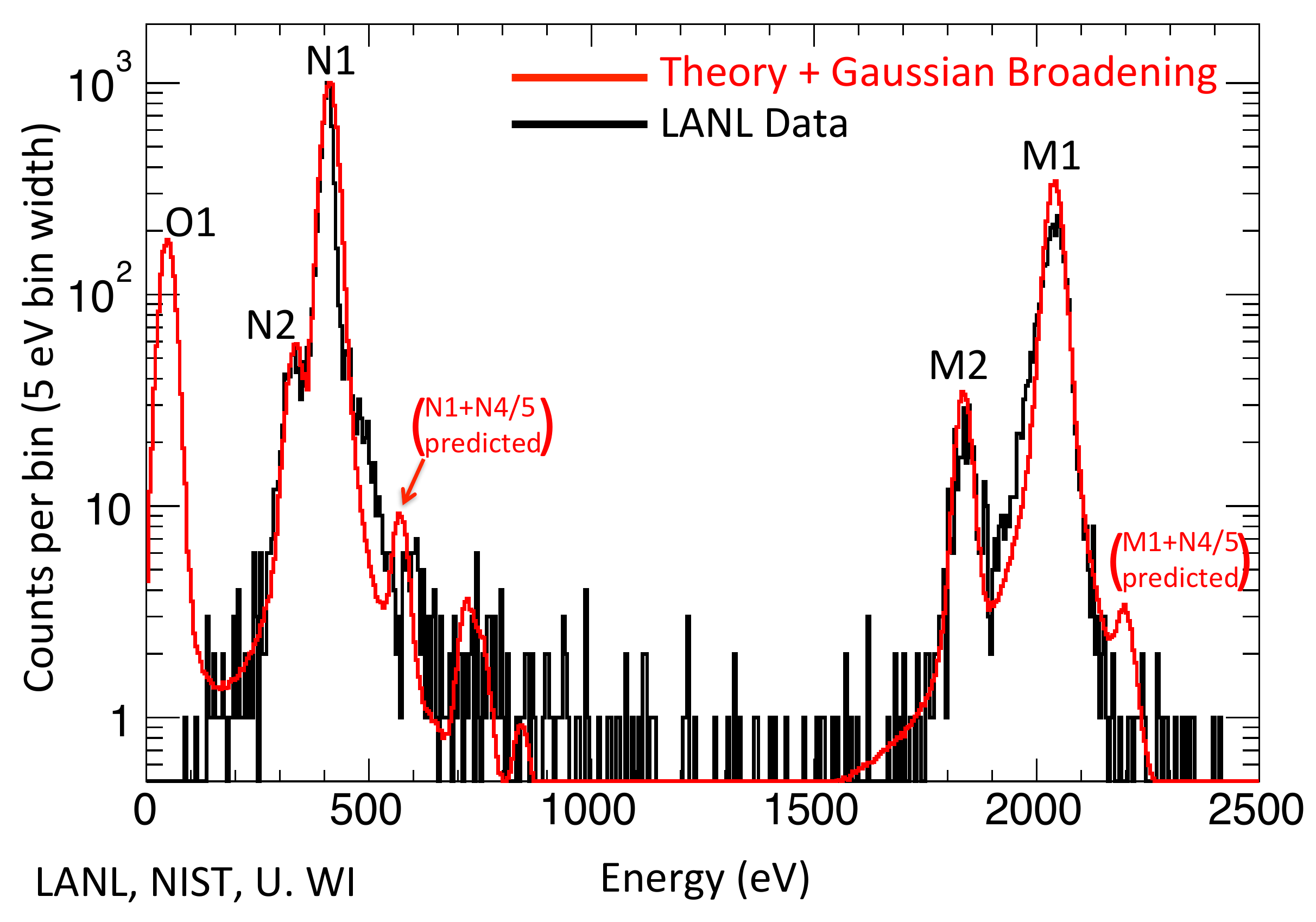}
\end{center}
\caption{Black:  calorimetric spectrum measured by NuMECS \cite{NUMECS2}.
Red: theoretical prediction of Faessler et al. \cite{Faessler2}. 
}
\label{fig:Kunde}
\end{figure}

Some of the preliminary data of ECHo \cite{ECHo2} and NuMECS \cite{NUMECS2} are 
shown in Figs.\,(\ref{fig:LG},\ref{fig:Kunde}). There, they are compared with an elaborate
theory  of the calorimetric spectrum  \cite{Faessler1,Faessler2},
 smeared with the experimental resolution 
and with the inclusion of one, two and three-hole
contributions. The agreement of theory and data is fairly good for the prominent M1, M2, N1 
and N2 contributions. 

There is in the ECHo data a significant peak at an
energy close to the expected position of a contribution from N1 capture accompanied
by O1 shakeup (N1O1), much larger than the theoretical expectation of
 \cite{Faessler1,Faessler2}. There is no evidence, at the predicted level, for 
the contribution from N1 capture accompanied by N4/5 shakeup (N1N4/5) in either 
experiment. An expected, sizeable M1N4/5 peak is also apparently absent in the NuMECS
data (it is tacitly assumed in these theoretical predictions \cite{Faessler1,Faessler2} 
that the computed probability for the production of a
second hole corresponds entirely to electron shakeup).
Finally, in both data sets, there is evidence for a
 ``shoulder'' above the theoretical expectations
in the 480 eV $< E_c <$ 550 eV domain of calorimetric energy.

\begin{figure}[htbp]
\begin{center}
\includegraphics[width=0.55\textwidth]{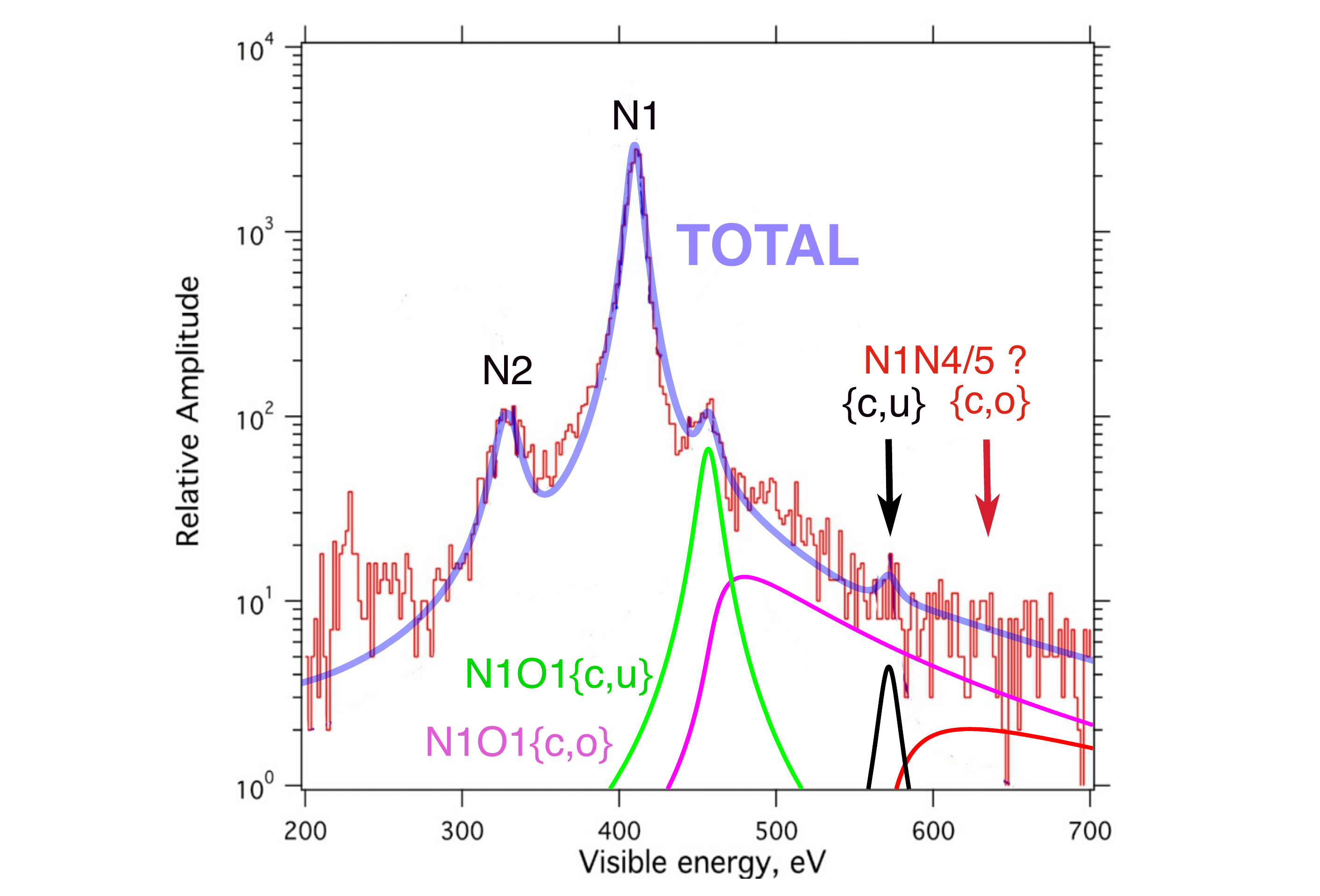}
\end{center}
\caption{The N-region. Slightly doctored predictions from \cite{us},
compared with early ECHo data, as
reported in \cite{Robertson1}. The excess below 300 eV is due to a $^{144}$Pm 
contamination. The resolution's width (FWHM) is taken
to be $\Delta E_{\rm exp}=8.4$ eV.
}
\label{fig:Us}
\end{figure}

In \cite{us} we have interpreted the situation just described in terms of a simple theory
of the production of ``second'' holes, described in the following chapter. The predictions,
for the region close to the N1, N2 single-hole peaks, are shown in Fig.\,(\ref{fig:Us}).
In this and subsequent figures, the symbols \{c,u\} and \{c,o\} refer to one electron
being captured (c), the other one being shaken up (u) or off (o). The curly brackets
remind us that the electrons are indistinguishable and their wave functions 
anti-symmetrized.

Once again,  theory  \cite{us}
 and data disagree. Our estimate of the height of the N1O1\{c,u\}
shakeup peak is
a factor $\sim\! 2.5$ too low. It is possible to correct in similarly moderate
ways the other contributions
such as to agree with the data.
One possibility, illustrated in Fig.\,(\ref{fig:Us}), is to correct the
N1O1\{c,u\} peak by the cited factor and
to leave N1O1\{c,o\} 
shake-off shoulder --which snugly
describes the observed one-- as predicted, while reducing the N1N4/5
features by a factor $\sim 3$.

\subsection{Lessons}

We conclude from the comparison of theory and preliminary data
that the predictions for the subdominant spectral features due 
to the production of more than one hole are to be taken {\it cum amplo grano salis}.

As we wall discuss in more detail in Sections VI and IX, it
is not a surprise
that the precise positions, heights and widths of the single- and multiple-hole
contributions ought to be fit to the observations.
Only the Breit-Wigner shape of a single peak (or shakeup multiple peak)
is theoretically hale and hearty. Much less so is
the shape of a shakeoff feature which, computed in the customary sudden approximation, 
depends on an overlap of bound and unbound wave functions.

Finally, the predictions for the normalization
of the various contributions, particularly the more elaborate ones \cite{Faessler1,Faessler2},
seem to be very untrustworthy. On the
way to an analysis of the endpoint,  the size of the
various spectral features will need to be adjusted to fit the data.
Currently, there are no theoretical results for the shake-off contributions, akin to those of  
\cite{Faessler1,Faessler2} for the shake-up probabilities. It would be useful to know whether
the shapes of these shake-off
contributions agree with those of our more naive approach which, as
we have seen, agree with the preliminary data.

\section{A rough theoretical guidance}
\label{theory}

In this section  
we clone the simple theory of the production of one \cite{ADRML}
or more holes \cite{us} in electron capture, adapting it to the region of calorimetric energies
extending from the M1 and M2 holes to the spectral endpoint.

\subsection{The spectrum of single holes}

In its simplest approximation \cite{ADRML} the spectrum of calorimetric energies, 
$E_c$, is a sum of Breit-Wigner (BW)
 peaks at the $E_c=E_{\rm H}$ positions\footnote{The daughter Dy* atom has a hole and an
 extra electron in the N7 ($4f_{7/2}$) level, of binding energy $\sim 4.2$ eV, so that, more
 precisely, $E_c=E_{\rm H}-E_{\rm N7}$. We do not in the text make this small
 correction. In comparisons with data the predictions for the peak positions
 will anyway be slightly adjusted to fit.}, 
 with their natural hole widths, $\Gamma_{\rm H}$. 
 The peak intensities are proportional to $\phi_{\rm H}^2(0)$, the values in Ho of the
 squared wave functions at the origin 
 of the electrons to be captured. The contribution of a single hole to the 
 EC decay rate, $R$, as a function of $E_c$ is: 
 \begin{eqnarray}
&& {dR[{\rm H}]\over dE_c}= \kappa\; E_\nu\,p_\nu\,n_{\rm H}\,
\phi_{\rm H}^2(0)\, BW[E_c,E_{\rm H},\Gamma_{\rm H}],
\label{dRdEc}\\
&& BW[E_c,E_{\rm H},\Gamma_{\rm H}]\equiv
{\Gamma_{\rm H}\over 2\pi}\,{1\over (E_c-E_{\rm H})^2+\Gamma_{\rm H}^2/4}\, ,
\label{BW}\\
&&E_\nu = (Q-E_c), \;\;\;\; p_\nu = \sqrt{(Q-E_c)^2-m_\nu^2}\, ,
\label{Eandp}
\end{eqnarray}
The factor $E_\nu$ in Eq.\,(\ref{dRdEc}) originates from the (squared) weak-interaction matrix element
and the factor $p_\nu$ in the decay's phase space. Having made explicit the $E_\nu$ factor,
$\kappa$ --in the excellent approximation in which nuclear recoil is neglected-- is a constant:
\begin{equation}
\kappa\;E_\nu \equiv {G_F^2\over 4\pi^2}\,\cos^2\theta_C\,B_{\rm H}\, |{\cal M}|^2,
\label{eq:resonances}
\end{equation}
with ${\cal M}$ the nuclear matrix element, $B_{\rm H}-1$
\cite{Bambynek}  an ${\cal O}(10\%)$ correction for atomic exchange and overlap
and
$n_{\rm H}$ the electron occupancy in the H shell of Ho 
(the actual fraction of the maximum number of electrons with the quantum numbers of H).

We know from the observations of neutrino oscillations that the electron
neutrino is, to a good approximation, a superposition of three mass 
eigenstates, $\nu_i$: $\nu_e=\sum_i U_{ei} \nu_i$, 
with $\sum_i |U_{ei}|^2= 1$. Thus, we ought to have written $dR[{\rm H}]/dE_c$
in Eqs.\,(\ref{dRdEc}-\ref{eq:resonances}) as an incoherent superposition 
of spectra with weights $|U_{ei}|^2$
and masses $m(\nu_i)$. But the measured differences 
$m^2(\nu_i)-m^2(\nu_j)$ are so small that 
the sensitivity of current direct attempts to measure $m_{\nu}$  
may result in a positive result 
only if neutrinos are nearly degenerate in mass, in which case $m_\nu$ in
Eqs.\,(\ref{dRdEc}-\ref{eq:resonances}) stands for their nearly common mass\footnote{If there were
 a small mixture of an extra mass eigenstate, it could be seen as a kink in the spectrum \cite{NOST}.}.

We have, in Eq.\,(\ref{dRdEc}), made the ``classical" approximation of neglecting interferences
between different vacated orbitals. For two different intermediate states H and H' to interfere, it
is necessary that they decay into the same subsequent state. The probability for this to happen,
in the case we are discussing, is small. An estimate of a generous upper limit to the effect of
interferences is discussed  in Appendix \ref{app:interferences}, 
demonstrating that interferences may be neglected.

\subsection{Shake-up}

As Robertson pointed out \cite{Robertson1,Robertson2}, in an EC
 event leading to a primary hole H
there is a small probability 
for a second hole H' to be made in a shake-up process. When the second electron is shaken-up
to any unoccupied daughter-atom bound-state level --of binding energy negligible relative
to $E_{\rm Tot}\sim E({\rm H})+E({\rm H'})$-- the calorimetric energy 
has a peak at $E_c\sim E_{\rm Tot}$. 

 To the extent that the
 presence of one hole does not significantly affect the filling --i.e.~decay-- of the other,
 the natural width of a two-hole state is the sum of the partial widths:
$\Gamma_{\rm Tot}\simeq\Gamma({\rm H})+\Gamma({\rm H'})$. In analogy
to the one-hole result of Eqs.\,(\ref{dRdEc}-\ref{eq:resonances}), the contribution
of a particular two-hole state to the calorimetric spectrum is:
\begin{eqnarray}
{dR[{\rm H,H'}]\over dE_c}&=&\kappa\; E_\nu\,p_\nu\,n_{\rm H}\,n_{\rm H'} \, BW[E_c,E_{\rm Tot},\Gamma_{\rm Tot}] \\
&\times& \sum_{n=7}^{\infty}
|\{1-\Pi({\rm H,H')}\}
\phi_{\rm H}(0)\,A{\rm (H,H',n)}|^2, \nonumber
\label{TwoHolesup}
\end{eqnarray}
where $n_{\rm H'}$ is the occupancy in the H' shell, $\phi_{\rm H}(0)$ is the wave 
function at the origin of the captured electron,
$A{\rm (H,H',n)}$ is the probability amplitude
for the excitation of the electron in H' to an unoccupied S-wave bound state with 
$n$ its principal quantum number\footnote{In the particular case of (H, H') = (M1, N1) one would have
$A{\rm (H,H',n)} = B_n$, as defined and estimated in Eq.(\ref{B}).},
and $\Pi({\rm H,H'})$ is the operator interchanging the two implicated electrons.
More precisely, $1-\Pi({\rm H,H'})$ stands for the operation of symmetrizing the
product of captured wave function  and transition amplitude in the singlet antisymmetric spin state,
antisymmetrizing it in the triplet symmetric spin state and
adding the resulting square moduli with weights 1/4 and 3/4.

\subsection{Shake-off}

The creation of a second hole H' in the capture leaving a hole H can also
occur as the shake-off of the electron in the orbital H' to the ``continuum''
of unbound electrons: 
\begin{equation}
{\rm Ho}\rightarrow {\rm Dy[{\rm H,H'}]}+e^-+\nu_e.
\label{eq:2holes}
\end{equation}
In such a 3-body decay, neither the electron nor the neutrino are
approximately monochromatic. 
The neutrino energy, $E_\nu$, and the ejected electron's kinetic energy, $T_e$, satisfy
$E_\nu+T_e=Q-E_{\rm Tot}$. The electron's energy and the daughter Dy ion 
energy excess add up to the observable calorimetric energy $E_c=T_e+E_{\rm Tot}$.

Let $|\rm Ho[H]\rangle$ be the wave function, in Ho, of the orbital whose electron is
to be captured and $\rm {| Dy[H,H'};p_e]\rangle$ the continuum wave function
of the electron ejected  off the daughter two-hole Dy ion. 
In the sudden limit
the shakeoff  distribution in electron momentum
$p_e$ (or in its energy $T_e$) ensues from the square of the wave function overlap:
\begin{eqnarray}
\!\!\!\!\!\!{dM\over dp_e}&\equiv&|\{1-\Pi({\rm H,H')}\}\,\phi_{\rm H}(0)\,
 \langle \rm {Ho[H'] | Dy[H,H'};p_e]\rangle|^2,
\label{overlapfreep}\\
\!\!\!\!\!\!{dM\over dT_e}&=&{m_e\over p_e}\,{dM\over dp_e},
\label{overlapfree}
\end{eqnarray}
where we have defined the auxiliary function $dM/dp_e$, to be used anon.

It is simplest to discuss the rate for the shake-off process of Eq.\,(\ref{eq:2holes})
by doing it for starters in the vanishing-width approximation for the daughter holes.
In this case
\begin{equation}
{dR\over dT_e}=\kappa\; E_\nu\,p_\nu\,n_{\rm H}\,n_{\rm H'}\,{p_e\over 4\pi^2}\,{dM\over dT_e}.
\label{edistr}
\end{equation}
The resulting $E_c$ distribution is:
\begin{equation}
{dR\over dE_c}=\int_0^{Q-E_{\rm Tot}}{dR\over dT_e}\,\delta(E_c-E_{\rm Tot}-T_e)\,dT_e\, .
\label{Ecdistr}
\end{equation}
To undo the zero-width approximation, substitute the above $\delta$ function by
$BW[E_c-T_e,E_{\rm Tot},\Gamma_{\rm Tot}]$, with $BW$ defined as in Eq.\,(\ref{BW}).

\subsection{The shake-off shape}
 
 Electron capture in Ho results in a Dy atom --which we shall
 in what follows denote as Dy*-- with a hole in the orbital
 from which the capture took place. The absent-electron
 charge partially shields the one of the absent-proton. Relative to a process without a similar
 effect --such as the creation of a primary hole by photo-ionization-- the partial shielding generally leads
 to a reduced probability for the creation of a second hole. This is because
 the wave functions of the
potentially vacated second orbitals in the parent and daughter atoms have a closer 
overlap in the presence of shielding. And --in the sudden approximation traditionally
used to make these kind of estimates-- the square of this overlap is the 
probability of 
not creating a second hole.

Intemann and Pollock \cite{I&P} were the first to show how to properly treat the 
shake-off of a second electron in EC. In what follows we 
apply their method and refer as an example to M1 capture accompanied by N1 shakeoff,
or, more precisely, to the M1N1\{c,o\} process (the most relevant one closest to the spectral endpoint).
The trick is to start by treating
the result of M1 capture (the absent proton and the absent electron)
 as a perturbation of the Coulomb potential of the form:
\begin{equation}
\alpha \, b(r)\equiv \alpha \left(
{1 \over r} - \int d^3r_1 \;{\vert\phi_{_{\rm M1}}(r_1)\vert^2\over \vert \vec r - \vec r_1\vert}\right).
\label{IP}
\end{equation}
 To first order in  $\alpha$ the wave function of the
N1 level in Dy* 
(the daughter atom with an M1 hole) 
is then expressed as a linear combination of Ho eigenfunctions: 
\begin{eqnarray}
|{\rm Dy^*[N1]}\rangle\simeq |{\rm Ho[N1]}\rangle+\sum_n B_n\,|{\rm Ho}[n]\rangle
\qquad\qquad\nonumber\\
+\frac{1}{2\,\pi}\int_0^\infty dp_e \; B_{\rm off}(p_e)\; |{\rm Ho}[p_e]\rangle,\qquad\qquad 
\label{wave}\\
B_n\equiv {\alpha \over E_{_{\rm N1}}-E_n}\int d^3r\,
\phi_n^* (r)\phi_{_{\rm N1}}(r)\, b(r),\qquad\qquad 
\label{B}\\
B_{\rm off}(p_e)\equiv{\alpha \over E_{_{\rm N1}}+T_e} \int d^3r\,\phi^* (p_e,r)\phi_{_{\rm N1}}(r)\, b(r)\;\quad
\label{BB}
\end{eqnarray}
where $n$ in the wave functions $|{\rm Ho[n]}\rangle$ stands for the $l=0$ bound levels with $n \neq 4$ and (positive) binding energy $E_n$, $\phi (p_e,r)$ is the unbound wave function
of the state $|{\rm Ho}[p_e]\rangle$,
 defined as in \cite{L&L}, of an $l=0$ electron with momentum $p_e$ and kinetic energy $T_e=p_e^2/(2\,m_e)$.

To take into account Fermi statistics, one must --it goes without saying--
 also do the calculations encapsulated in 
Eqs.\,(\ref{IP}-\ref{BB}) with the exchange M1 $\leftrightarrow$ N1, since the two-hole final state,
at a given $T_e$, is the same independently of which electron is captured or ejected. 
Thus, in the sudden approximation, the
square of the (monopolar) matrix element
for an electron being shaken-off with energy $T_e$ results in:
\begin{equation}
{dM\over dT_e}= {m_e\over 4\,\pi^2\,p_e}\,|\{1-\Pi({\rm M1,N1)}\}\,\phi_{\rm M1}(0)
B_{\rm off}(p_e)|^2
\label{antiSym}
\end{equation}
for the distribution function of Eq.\,(\ref{overlapfree}).

A crucial point in estimating wave-function overlaps is to choose them with the correct
spatial scale. To provide an estimate of the shape of $dM/dT_e$ we shall use 
non-relativistic Coulomb 
wave functions\footnote{These would give very poor estimates of the wave functions 
at the origin, for which we use instead 
the values given by the accurate calculations in \cite{Robertson2, Faessler1, Faessler2}.}  of Ho
with effective values of $Z$ chosen to reproduce the relevant
orbits' mean radii, as calculated with more precise Hartree-Fock methods \cite{McLean}.
Let $r_B\equiv 1/(\alpha \, m_e)$ denote the Bohr radius. 
For $\langle r({\rm M1})\rangle=0.246\,r_B$ and
$\langle r({\rm N1})\rangle=0.555\,r$ 
the effective charges are $Z_{\rm eff}({\rm M1})=54.9$, to be used in Eq.\,(\ref{IP}) and
$Z_{\rm eff}({\rm N1})=43.2$ , to be used for the bound and free wave functions
in Eqs.\,(\ref{B},\ref{BB}).

\section{The M-hole region and beyond}

In what follows we present results for the spectral domain extending from the M2 and M1
single-hole contributions to the spectral endpoint, assumed to be at Q = 2833 keV. First
we take at face value the results of the calculations described in the previous section.
Since the results are very optimistic --in the sense of facilitating a potential constraint on
$m_\nu$-- we shall later discuss the possibility that our predictions are gross overestimates.

\begin{figure}[htbp]
\begin{center}
\includegraphics[width=0.48\textwidth]{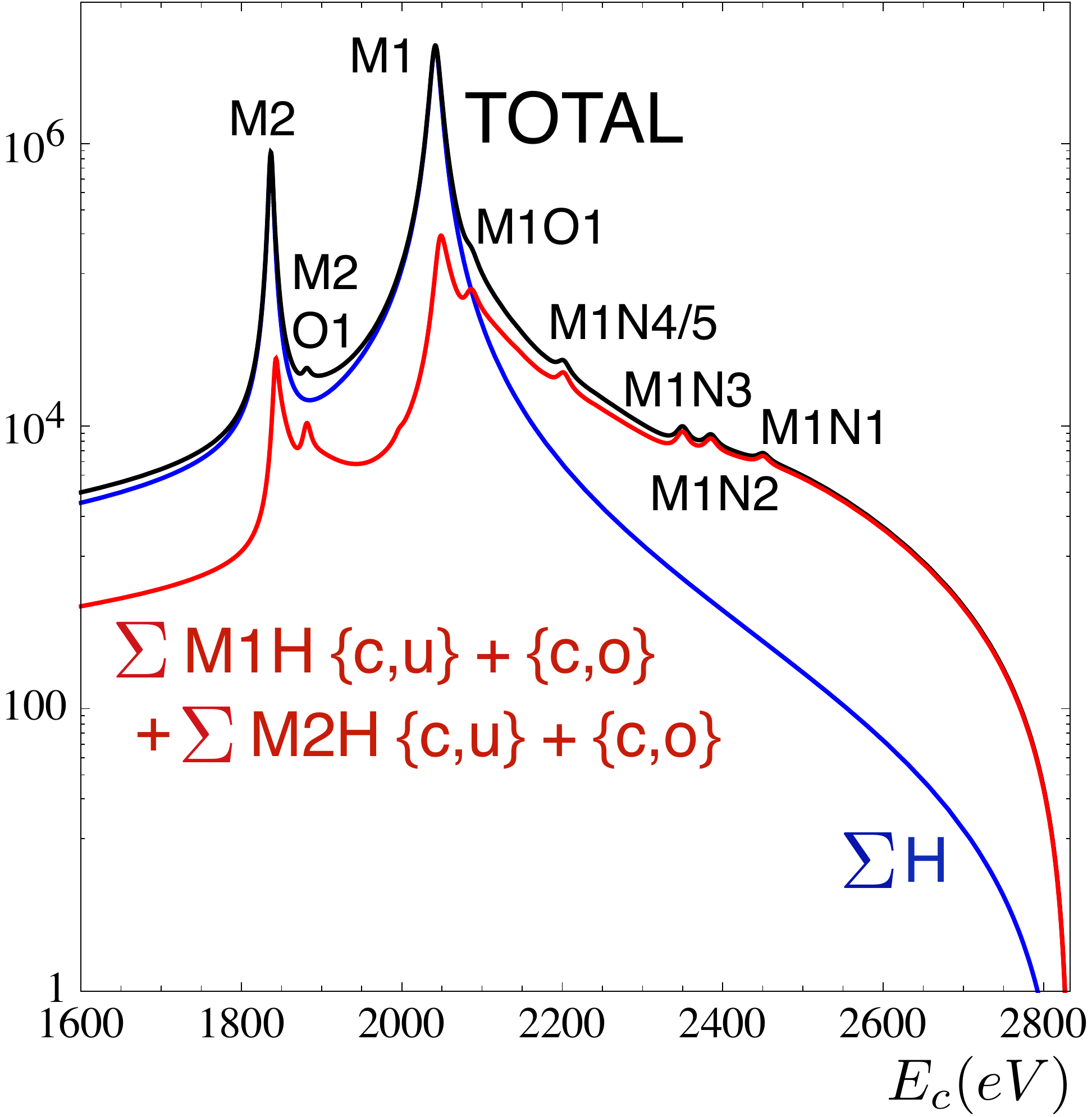}
\end{center}
\caption{The ``theorist's" calorimetric spectrum in the M region. 
In blue the single-hole contribution $\Sigma\rm H$. In red the two-hole shake-up plus shake-off contributions, with one of the holes being M1 or M2. In black, the sum of all contributions. 
The units of the vertical scale are arbitrary.
}
\label{fig:FullMSpectrum}
\end{figure}

In Fig.\,(\ref{fig:FullMSpectrum}) we show the separate contributions of the
one-hole spectrum, the dominant contribution to the two-hole spectrum 
in this energy domain (M2H plus M1H), and the one-plus-two-hole 
result\footnote {We did not include other two-hole contributions (like N1M4 and N1M5) 
giving small structures below the M2 peak.}.
This is a {\it theorist's spectrum}, 
with an arbitrarily normalized vertical scale,
$m_\nu=0$ and an experimental width, $\Delta E_{\rm exp}$,
whose square is negligible relative to that of any of the hole's natural widths.
All the subsequent figures will also, in the same sense, depict 
arbitrarily normalized spectral shapes,
occasionally with $m_\nu\neq 0$.

\begin{figure}[htbp]
\begin{center}
\includegraphics[width=0.5\textwidth]{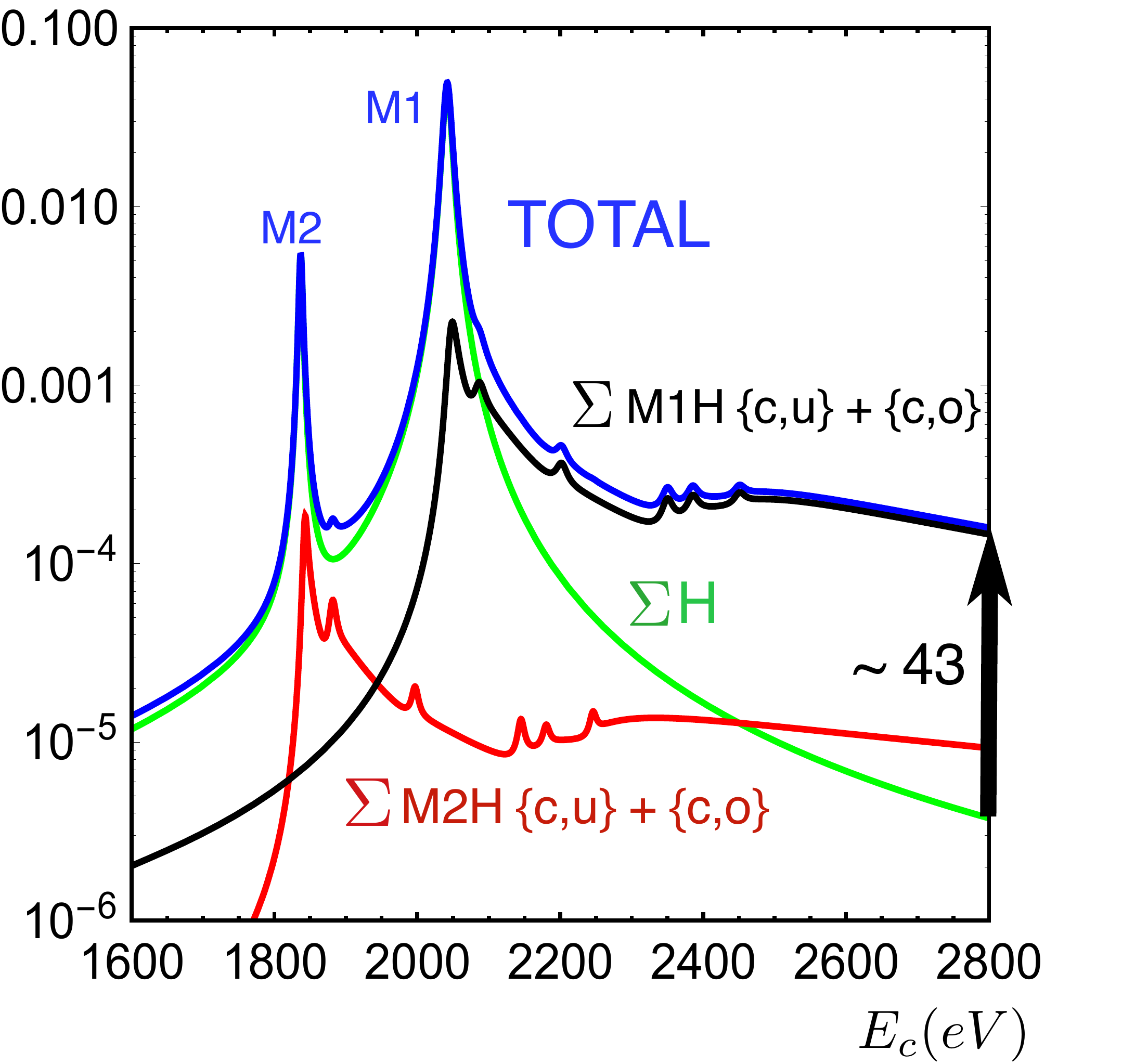}
\end{center}
\caption{The ``bare" spectrum defined in the text. 
The normalization of the ordinate is arbitrary. In black (red) the sum of the two-hole spectrum, 
one of the holes being M1 (M2). In green, the single-hole contribution. In dark blue, the total. 
The units of the vertical scale are arbitrary.}
\label{fig:M12HHoles}
\end{figure}

The enhancement above the M1 peak of the total spectrum in Fig.\,(\ref{fig:FullMSpectrum}), 
relative to the single-hole result, is quite considerable. A very good way to present this phenomenon is to plot the {\it bare spectrum}, i.e.\,the result of dividing the numbers of events of
 Fig.\,(\ref{fig:FullMSpectrum}), by $E_\nu\,p_\nu$, the
factors originating from the nuclear matrix element and the phase space, respectively.
This is done in Fig.\,(\ref{fig:M12HHoles}). In this and subsequent plots of ``bare spectra"  the
vertical scale is arbitrary and we do not even label it.

Notice in Fig.\,(\ref{fig:M12HHoles}) that, at $E_c \sim \rm Q$, the two-hole contributions are overwhelmingly dominant, a factor $\sim 40$ larger than the single-hole 
contribution\footnote{The figure is plotted for a fixed height of the M1 peak. 
With this constraint, 
the NH and MH contributions increase the overall normalization of the spectrum by $\sim 9$\%,
so that, for a fixed total number of events, the enhancement would be closer to 40 than to 43,
should we trust the double-hole predictions to this level of precision.}. 
In a closer
look at the figure one concludes that this large effect would be equivalent, in the absence of 
two-hole contributions, to a Q-value of $\sim 2150$ eV, a point of the green single-hole
curve with an ordinate as high as the endpoint of the (total) blue curve.

\begin{figure}[htbp]
\begin{center}
\includegraphics[width=0.48\textwidth]{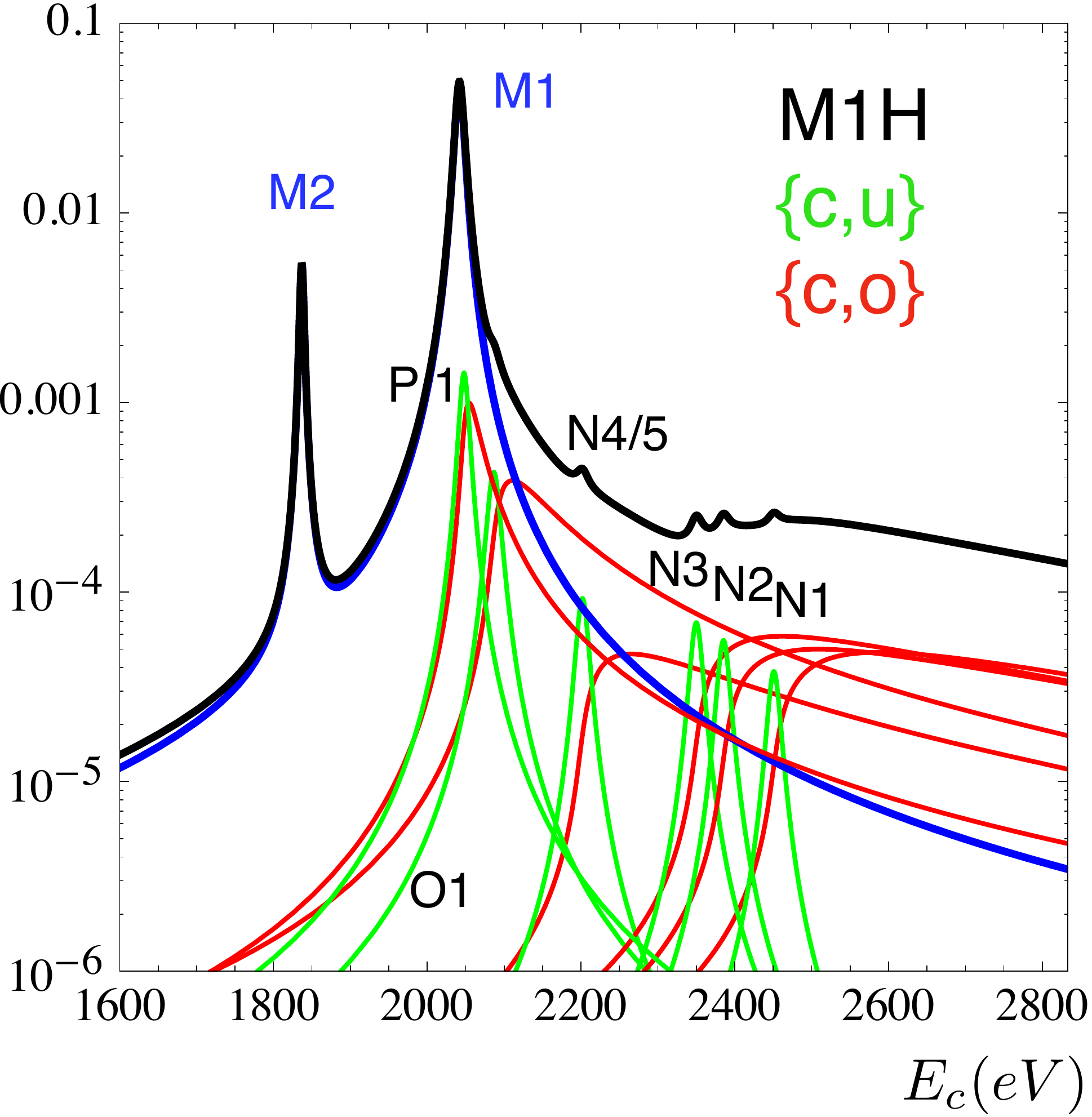}
\end{center}
\caption{The M1H two-hole contributions in Fig.\,(\ref{fig:M12HHoles}), in full detail. 
In green (red), one of the holes 
is captured, the other one shaken up (off). In black a total bare spectral shape, including the (blue) single-hole contribution, but not the M2H contributions, negligible above the M1 peak.
The units of the vertical scale are arbitrary.
}
\label{fig:M1HHoles}
\end{figure}

An analogous improvement is expected on another subject: 
the possible observation of antineutrinos of the cosmic background via their capture in $^{163}$Ho. The quantity of radioactive isotope necessary to obtain ten events of signal is a strongly increasing function of the Q-value  \cite{LusVig}: with only single-hole EC it should be 1274 (23.2) kg y for 
Q = 2.8 (2.2) keV. Including the contribution to the two-hole spectrum the same quantity comes out to be 30.6 kg y even if Q = 2.833 keV. Gathering this amount of $^{163}$Ho is still a wee bit unrealistic. 

\begin{figure}[htbp]
\begin{center}
\includegraphics[width=0.48\textwidth]{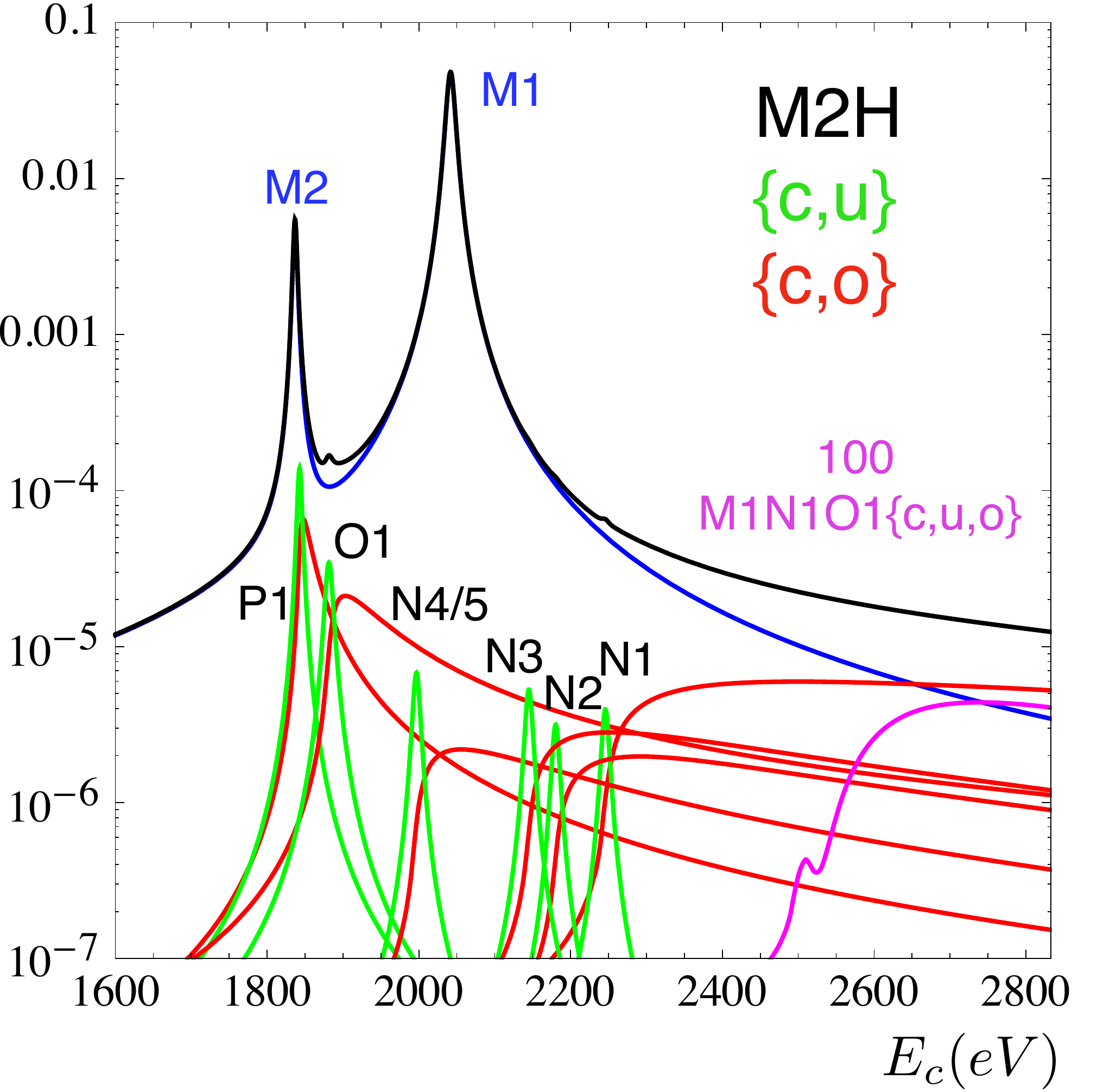}
\end{center}
\caption{The M2H two-hole contributions in detail. In green (red), one of the holes 
is captured, the other one shaken up (off). In black a total bare spectrum including the (blue) single-hole contribution, but --unrealistically-- not the M1H contribution. Enhanced by two orders of magnitude, the M1N1O1\{c,u,o\} three-hole contribution is also shown. 
The units of the vertical scale are arbitrary.
}
\label{fig:M2HHoles}
\end{figure}

The details contributing to the construction of Fig.\,(\ref{fig:M12HHoles}) are given
in Figs.\,(\ref{fig:M1HHoles},\ref{fig:M2HHoles}), where, respectively, the various
M1H and M2H double-hole contributions are specified.

We have seen that some two-hole effects successfully compete with single-hole ones when the sizable contribution 
of the former is at $E_c$ values at which the latter is not close to its peak. The triple-hole 
contributions, contrariwise, are largest at $E_c$ values at which the shake-off contribution
of double holes is relatively large. Given the smallness of the extra wave-function overlap 
in three-hole contributions (relative to the two-hole ones), three-hole effects are always negligible.
The example of M1N1O1\{c,u,o\}, arbitrarily multiplied by a factor of one hundred,
 is shown in Fig.\,(\ref{fig:M2HHoles}). 

\section{The endpoint analysis}

\begin{figure}[htbp]
\begin{center}
\includegraphics[width=0.48\textwidth]{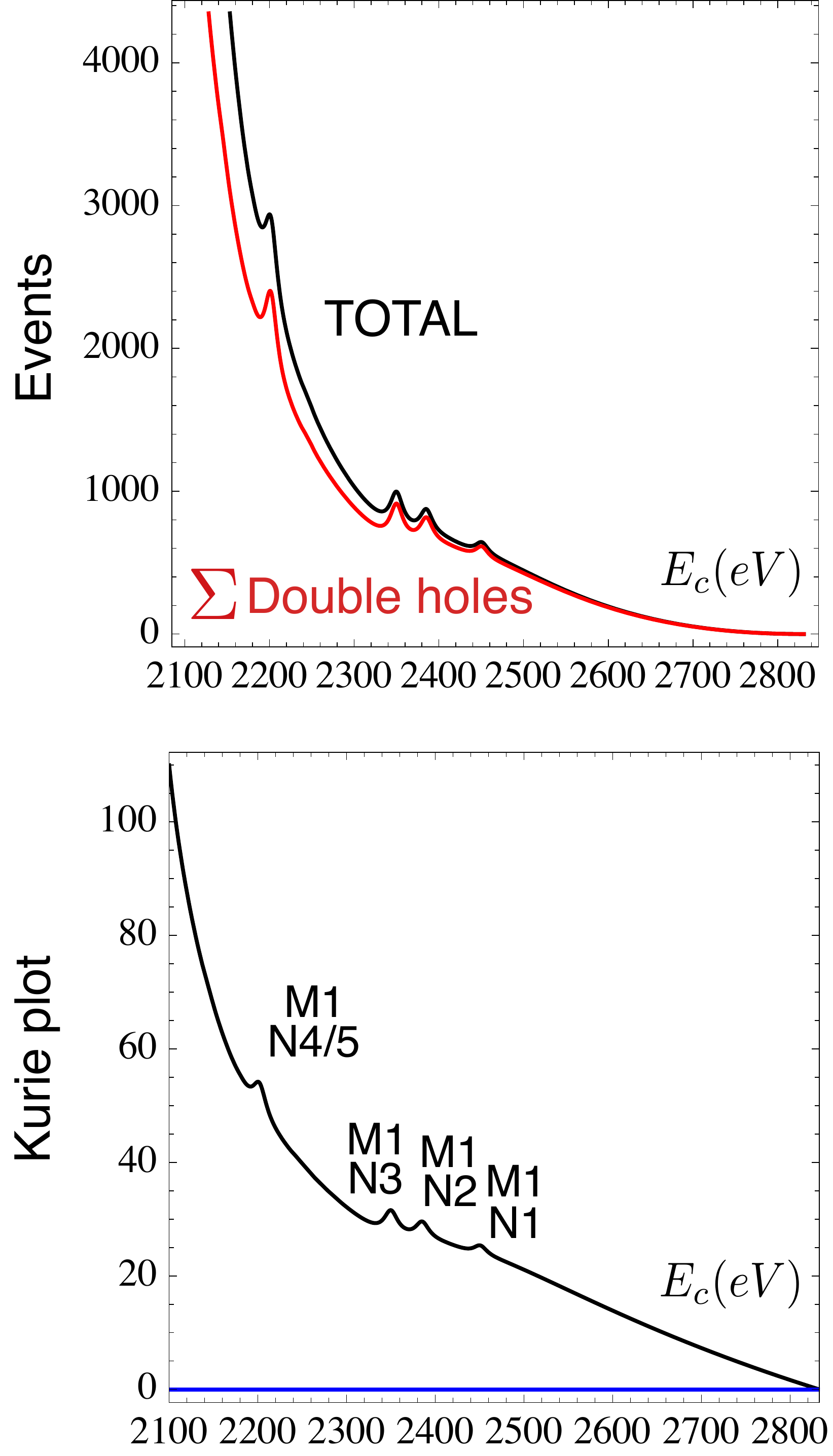}
\end{center}
\caption{{\bf Above}: The uppermost $\sim 700$ eV of the spectrum, with an arbitrary
vertical scale. {\bf Below}: The corresponding Kurie plot (the square root of above-plotted
spectral shape).
}
\label{fig:KurieA}
\end{figure}

The spectral shape in the last $\sim 700$ eV below the endpoint is shown in the upper
 Fig.\,(\ref{fig:KurieA}), where the dominance of double-hole contributions is predicted
 to be most significant. The lower figure is a ``Kurie"
 (or Kurie-like plot), simply depicting the square root
 of the number of (theorist's) ``events". Notice that in the last 200 eV,
 there being no significant spectral features (double hole \{c,u\} peaks or
 \{c,o\} thresholds) the Kurie plot is 
 linear  but, as we proceed to discuss, not quite. Recall here, and in what follows, that
 we have assumed that our ``experiment" has a perfect resolution.

\begin{figure}[htbp]
\begin{center}
\includegraphics[width=0.48\textwidth]{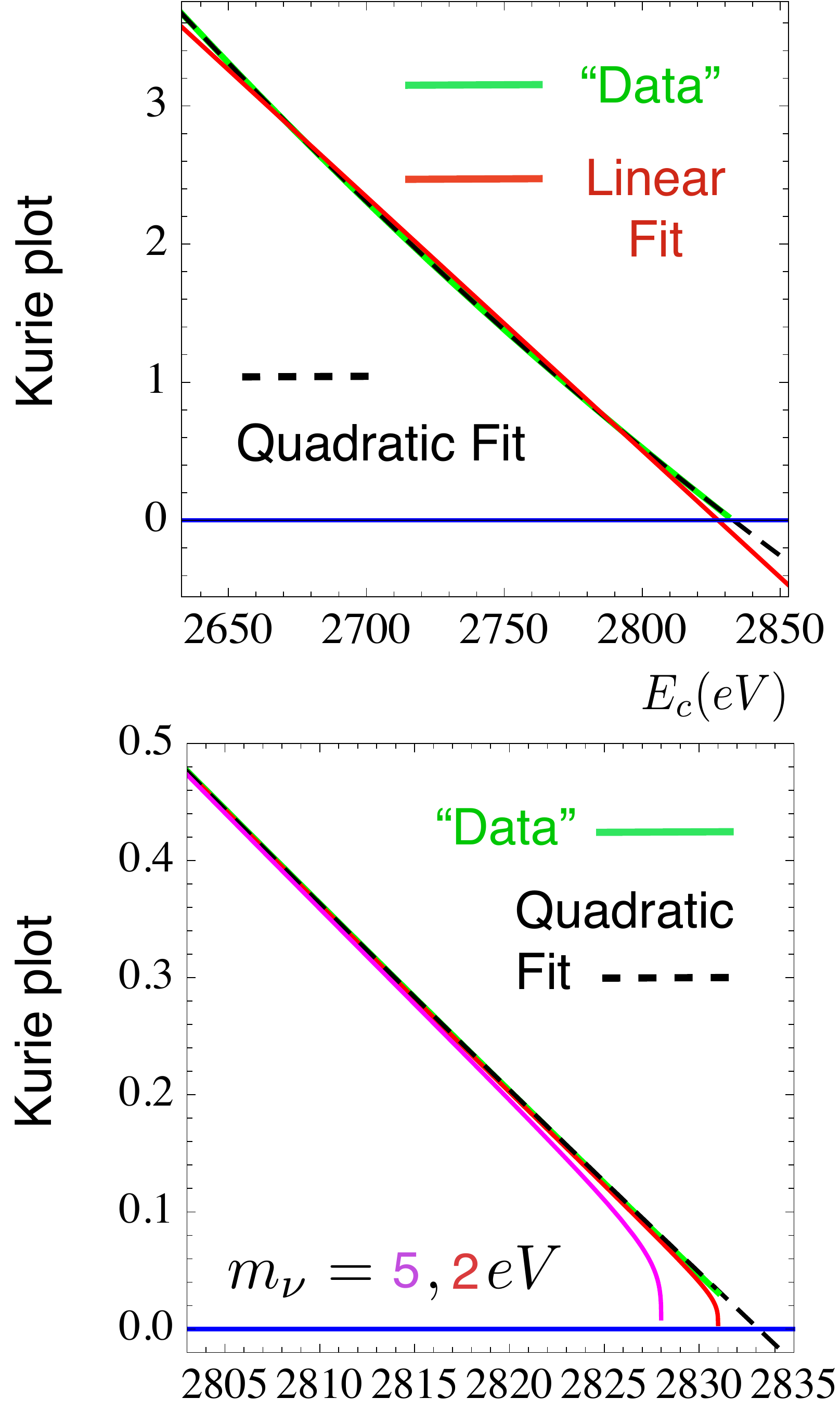}
\end{center}
\caption{{\bf Above}: The uppermost $\sim 200$ eV of the Kurie plot,
showing a linear and a quadratic fit to the theoretical Q = 2833 eV,
$m_\nu=0$  ``data".
{\bf Below}: The last $\sim 30$ eV, including the expectations
for $m_\nu=2,\,5$ eV, which disagree with the input ``data".}
\label{fig:KurieB}
\end{figure}

The uppermost $\sim 200$ (30) eV of the Kurie plot are shown in 
the upper (lower) part of Fig.\,(\ref{fig:KurieB}). We are only interested
in checking the degree of non-linearity of this theoretical plot, so our disrespect
for statistical issues will be irrelevant. Our theorists' ``Kurie data",
generated with $m_\nu=0$, Q = 2833 eV,
are fit to a linear and a quadratic polynomial in $E_c$, in a theorist's
way. To wit, the interval ${\rm Q}-200\le E_c\le \rm Q-2$ eV (or Q - 10 eV, to check
that the a-priori ignorance of a precise calorimetric Q-value makes little difference) 
is binned in $\sim 100$ 2-eV intervals. Two least
square fits to this binned ``data" are performed, ignoring the fact that, in reality,
the real data would be statistically more precise as $E_c$ decreases.

The results of the fits are:
\begin{eqnarray}
\!\!\!\!\!\!\!\!\!{\rm A}(E_c) &=& 51.9226 - 0.018363 E_c, \label{line}\\
\!\!\!\!\!\!\!\!\!{\rm B}(E_c) &=&  162.652 - 0.0994605 E_c + 1.4842 \,10^{-5} E_c^2. \label{parabola} 
\end{eqnarray}

The linear fit is not very good, while the quadratic one very snugly describes the ``data".
The condition ${\rm A}(E_c) =0$ results in Q = 2827.56 eV, wrong by $\sim 0.19$\%
relative to the ``data's" Q = 2833 eV. The condition ${\rm B}(E_c) =0$ results in 
Q = 2833.14 eV, correct to 5 parts in $10^5$.

 In the lower part of Fig.\,(\ref{fig:KurieB}) we test ``by eye" how the mentioned
 ``Kurie data" or the linear and quadratic fits thereof would, given the required
 statistics and the other obvious provisos, exclude neutrino masses of 2 or 5 eV.

\section{An evident caveat}

We have been treating our predictions as if they were to be precisely trusted.
But we have seen, by comparing them with the preliminary data in the N-region of energies,
that they are not. The theory appears to overestimate some double hole contributions,
and to underestimate others. Moreover
our results about the analysis of the endpoint appear
to be unprecedently optimistic, and this will be even more so when we discuss the pile-up issue.
Thus, we adopt in this section a complementary very pessimistic point of view.

\begin{figure}[htbp]
\begin{center}
\includegraphics[width=0.48\textwidth]{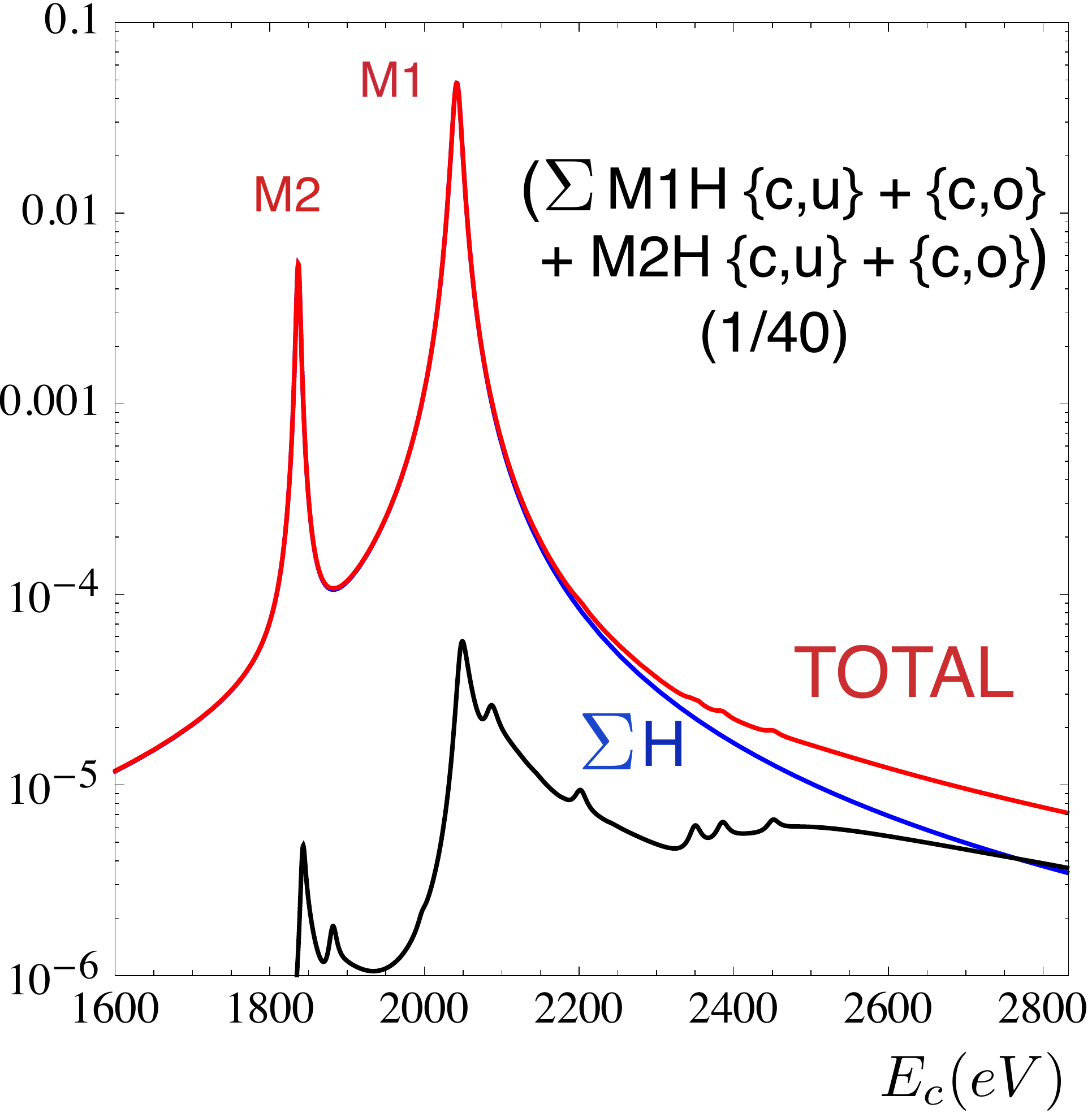}
\end{center}
\caption{Bare spectrum with all double-hole contributions arbitrarily
divided by 40, with their sum in black.
Blue: the sum of single holes. Red: total.
}
\label{fig:BareOver40}
\end{figure}

Assume that all the double hole contributions relevant to the endpoint analysis 
(M1H and M2H) have been overestimated by a collective factor of 40, chosen to have their
sum at $E_c\sim \rm Q$ be comparable to the M1-dominated single hole
contribution in that domain. The corresponding bare spectral shapes are
shown in Fig.\,(\ref{fig:BareOver40}).

\begin{figure}[htbp]
\begin{center}
\includegraphics[width=0.48\textwidth]{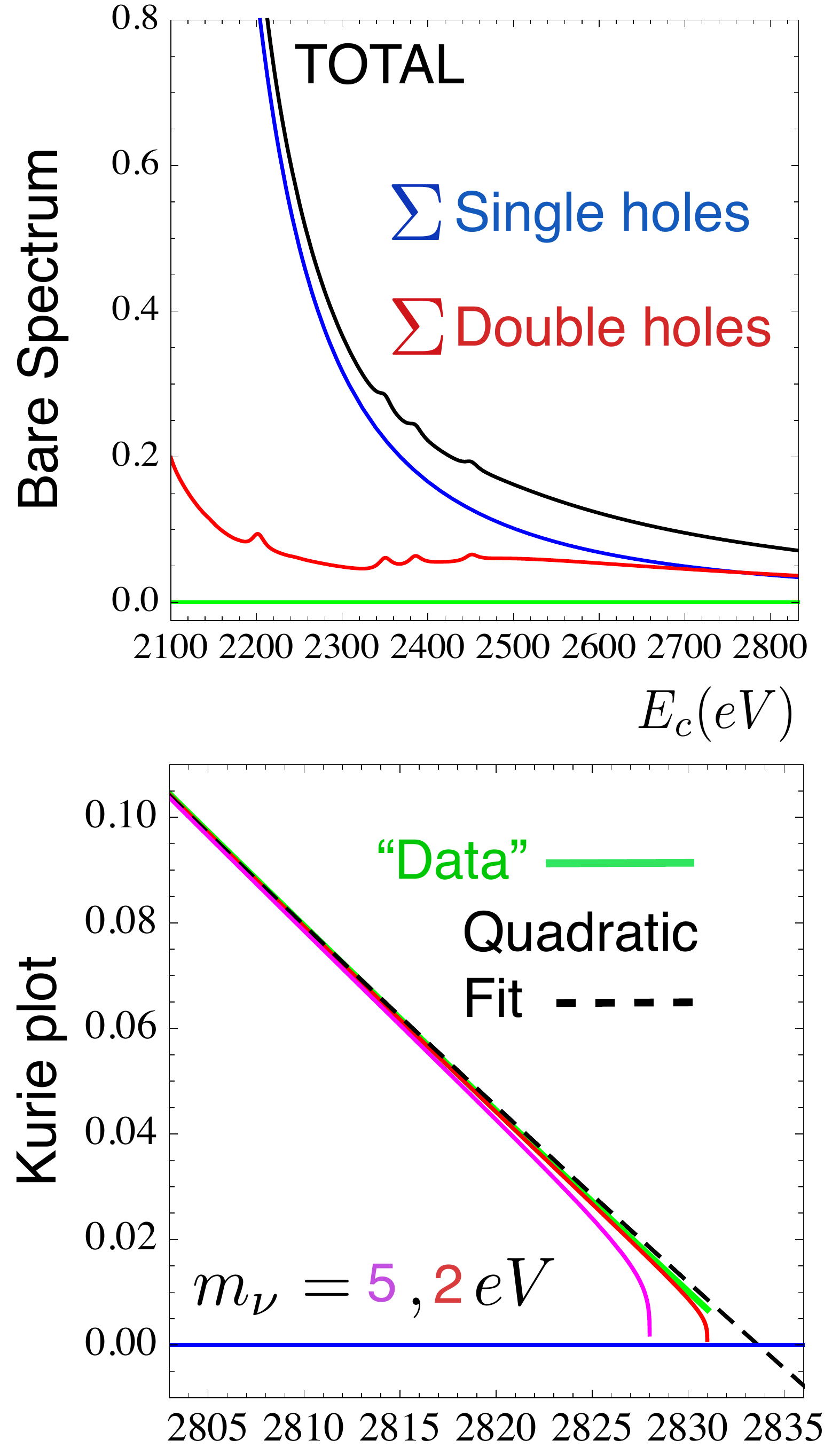}
\end{center}
\caption{{\bf Above}: Bare spectrum of the uppermost $\sim 700$ eV of an $m_\nu=0$  
set of ``data" with the
predicted two-hole contributions arbitrarily diminished by a factor of 40. 
{\bf Below}: The corresponding Kurie plot in the last 30 eV, showing also 
a linear and a quadratic fit, as well the theoretically excludable expectations
for $m_\nu=2,\,5$ eV.
}
\label{fig:KurieOver40}
\end{figure}

In the upper part of Fig.\,(\ref{fig:KurieOver40}) we show the bare spectrum, at energies
 above $\rm Q - 700$ eV. The lower part is the corresponding Kurie plot, showing
 a quadratic fit to the  ``data" analogous to that described in connection with
 the lower part of Fig.\,(\ref{fig:KurieB}).
 
 Once again, we make linear and quadratic fits to the ``data", binned in the last $\sim 200$ eV
 below the endpoint. The results are:
\begin{eqnarray}
\!\!\!\!\!\!\!\!\!{\rm A'}(E_c) \! &=& \! 11.9492 \! - \! 0.0042288\, E_c, \label{line2}\\
\!\!\!\!\!\!\!\!\!{\rm B'}(E_c) \! &=& \! 47.8908 \!-\! 0.0305521\, E_c \!+\! 4.81759\, 10^{-6}\, E_c^2. \label{parabola2}
\end{eqnarray}

The linear fit is not very good, while the quadratic one very snugly describes the ``data".
The condition ${\rm A'}(E_c) =0$ results in Q = 2825.67 eV, wrong by $\sim 0.26$\%
relative to the ``data's" Q = 2833 eV. The condition ${\rm B'}(E_c) =0$ results in 
Q = 2833.65 eV, correct to 2.3 parts in $10^4$.

\subsection{A preliminary conclusion}

Assume that, in a particular experiment,
 the experimental resolution function and the background are well understood.
Traditionally, Kurie plots for well understood processes --such as tritium $\beta$-decay--
are defined in such a way that they are expected to be linear in energy.
They are then fit with three parameters:
$m_\nu$ and the constant and the slope of a linear function of energy
or, equivalently, $m_\nu$, the slope and Q.

In the case of the calorimetric measurement in $^{163}$Ho EC the
shape of the BW functions describing single-hole or HH'\{c,u\} double-hole
contributions, as we shall discuss, are very well understood. The ``problem" is that
the precise shape and magnitude of  the HH'\{c,o\} double-hole contributions are not. 
Even if these contributions are measured to be significant as one approaches
the endpoint, we have argued, the use of one more parameter in the fits (the extra 
coefficient in a quadratic function of energy) should solve this apparent problem.

It goes without saying that these theoretical conclusions would gain (or lose) 
weight if and when they are tested with adequate simulations of data in a
realistic observational environment. The advantages of a possibly increased
statistics in the endpoint region (compared with single-hole expectations) 
may be a tempting reason to perform such an analysis.

\section{Pile-up}

The finite time required to record an EC event gives rise to the
inescapable problem of pile-up: the additional contribution of the 
spectrum of two ``simultaneous" events. If the ``singles" spectrum (of single events)
has a sizable contribution of HH'\{c,o\} events towards its endpoint, the
concern with pile-up {\it diminishes}.

Let $d\hat R/dE_c$ be the singles spectrum, with its integral in the
interval  $0<E_c<Q$ normalized to unity. 
The pileup spectrum, also normalized to unity in the interval  $0<E_c<2\,Q$, is:
\begin{equation}
{dR_P\over dE_c}=\int_0^Q \! dE_1\int_0^Q \! dE_2\, {d\hat R\over dE_1}\,{d\hat R\over dE_2}
\,\delta(E_c-E_1-E_2).
\label{pileup}
\end{equation}

\begin{figure}[htbp]
\begin{center}
\includegraphics[width=0.48\textwidth]{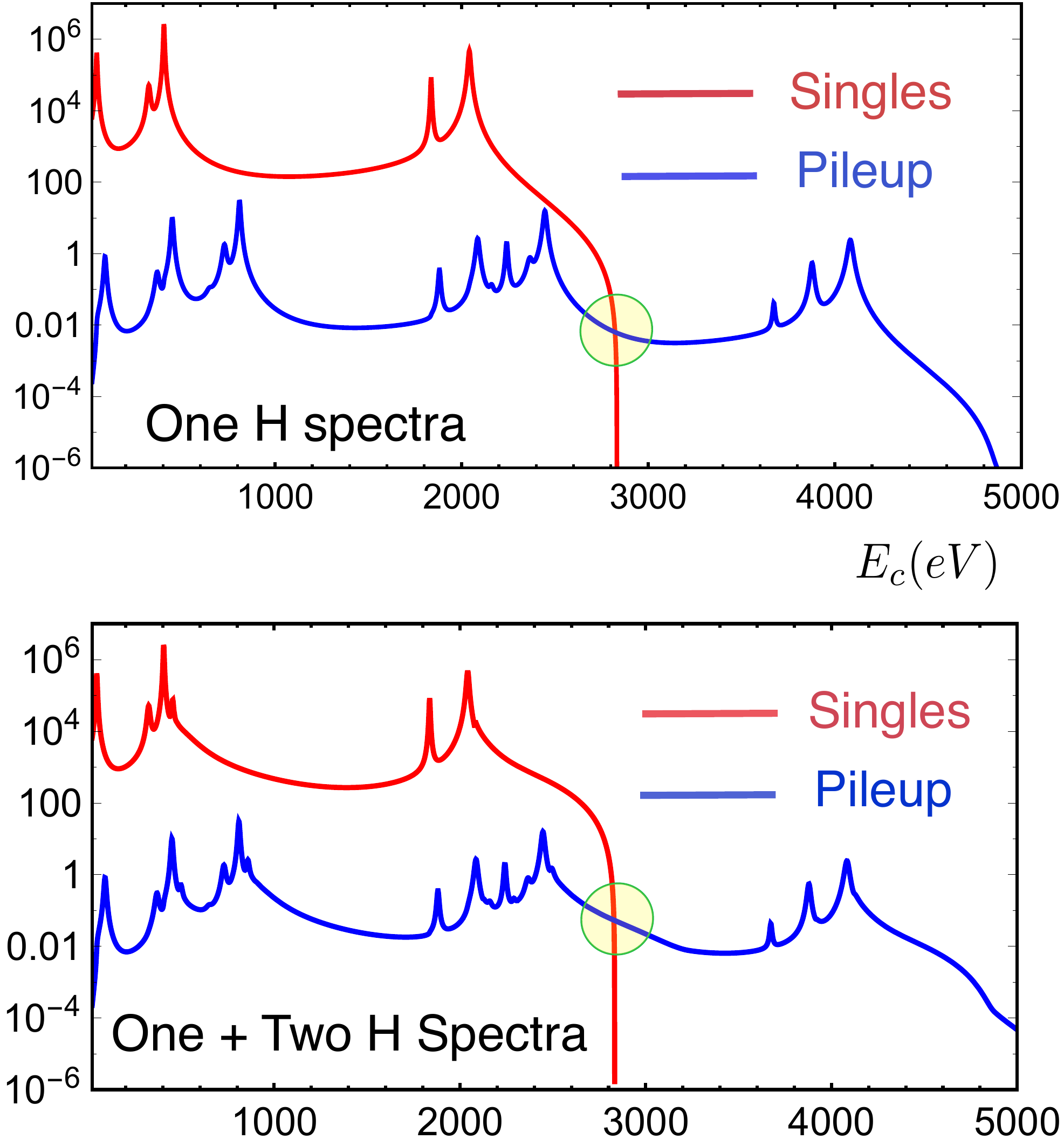}
\end{center}
\caption{``Singles" 
and ``Pileup" 
spectral shapes, with a $4\times 10^{-5}$ pileup probability
and an arbitrary (but common) vertical scale normalization.
{\bf Above}: Spectra with captures resulting only in single holes.
{\bf Below}: Spectra with the addition of events resulting in double holes.
The regions around $E_c=Q$ are significantly
different in the upper and lower plots.
}
\label{fig:AllPileup}
\end{figure}

For the sake of illustration, let the probability of pileup be $4\times 10^{-5}$.
In the upper part of Fig.\,(\ref{fig:AllPileup}) we show the corresponding
singles and pileup spectra for the case in which they are exclusively dominated
by single-hole captures; the vertical scale is arbitrary, but the ratio of the
two spectra is not, it corresponds to the assumed 
energy-integrated
pileup probability.
With the same proviso\footnote{A hawk-eyed reader may notice
that these spectra do not have the little wiggles visible in 
Fig.\,(\ref{fig:FullMSpectrum}). They have been smoothed in the benefit
of a faster numerical integration of Eq.\,(\ref{pileup}).}, 
the spectra corresponding to our estimate 
of two-hole effects are shown in the lower part of Fig.\,(\ref{fig:AllPileup}).

\begin{figure}[htbp]
\begin{center}
\includegraphics[width=0.48\textwidth]{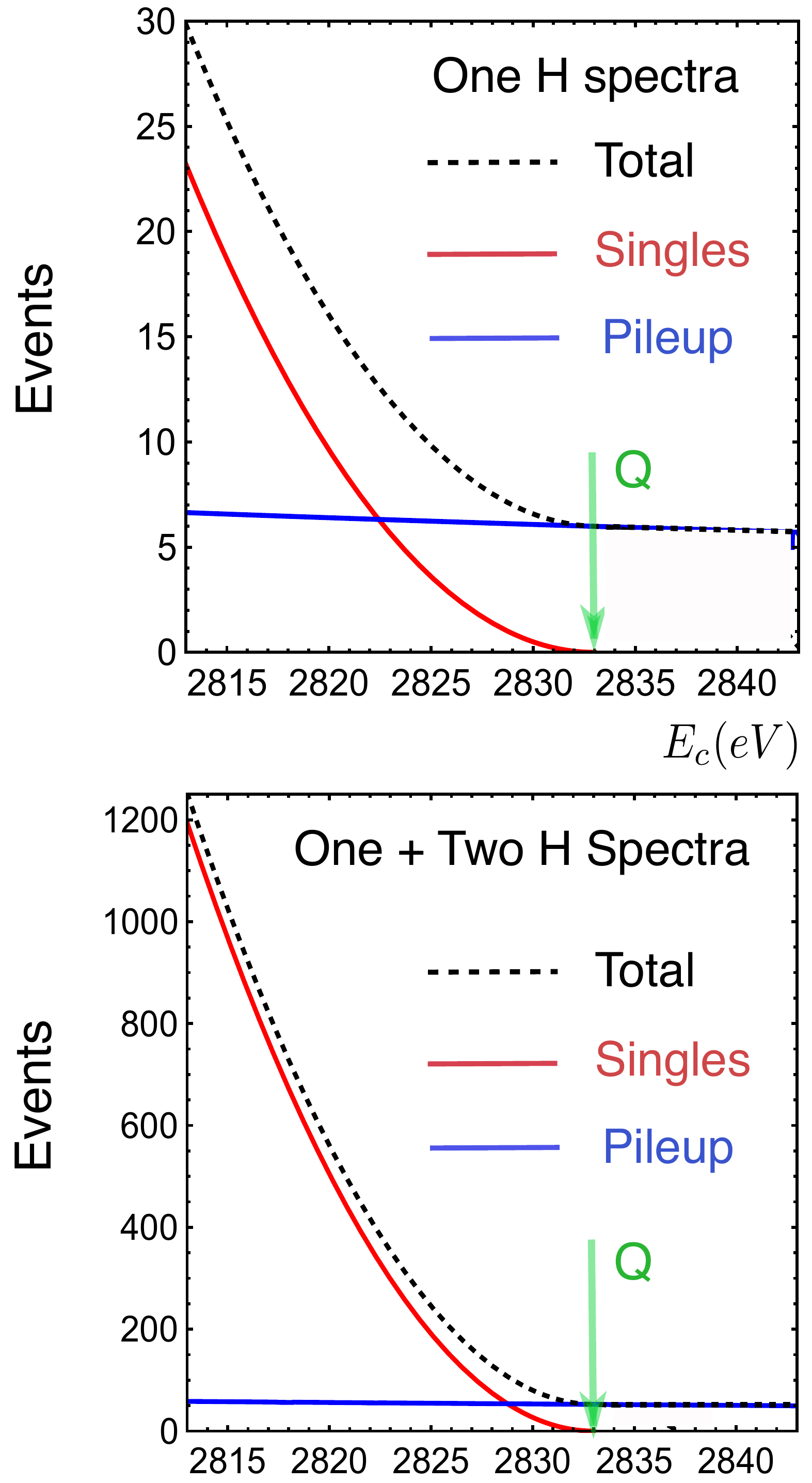}
\end{center}
\caption{ Blowup of the domains $Q -20\,{\rm eV}<E_c<Q+10$ eV of Fig.\,(\ref{fig:AllPileup}).
{\bf Above}: Spectrum dominated by single holes.
{\bf Below}: With the addition of two-hole processes.
The vertical scales of the two plots correspond to the same assumed activity of the source
and data-taking time.
}
\label{fig:EndpointPileup}
\end{figure}

Comparing the emphasized regions in the upper and lower parts of Fig.\,(\ref{fig:AllPileup})
one concludes that the end-point pileup problem is much less serious in the case that
the double-hole contributions are significant there. This is made clearer by comparing
the $E_c\sim \rm Q$ domains, which we do in Fig.\,(\ref{fig:EndpointPileup}).
The ratio of single to pileup events at, for instance, $E_c=\rm Q -3$ eV is $\sim 6$
times larger in the lower part of this figure (where two-hole effects are included) than
in the upper part (where they are not).

The reason why we find improved expectations concerning pileup is simple. At
$E_c\sim$ Q the pileup spectrum is dominated by the addition of the N1H\{c,o\} and M1H\{c,o\}
spectral tails, while the singles spectrum is dominated by the latter. 
But  in the pile-up integral of Eq.\,(\ref{pileup}) all spectral features are 
partially smoothed out.

\section{Single-hole peaks, BW tails and residua}

A clear conclusion from the existing data is that there are spectral contributions not
anticipated in a simple single-hole theory. In the current exploratory phase 
of the experiments it would be tempting to subtract from the data the single-hole
expectation, to visualize directly the {\it residuum}: the cited extra contributions. 
One could, for instance,
subtract the blue curve in Fig.\,(\ref{fig:FullMSpectrum}) from the black one, to
extract the red one: the residual sum of double-hole contributions. We now address the
question of the precision with which this  can be done. 

Given the large statistics with which single-hole contributions will be measured at
and around their peaks, there is no question that their individual parameters 
(position, height and width) will be exquisitely measured. But to obtain the residua
a BW shape in Eqs.\,(\ref{dRdEc},\ref{BW}),
extending up to very many widths above the peak, must be
assumed. A Breit-Wigner is an approximation known to require, in certain cases,
potentially large explicit corrections \cite{PDG}.

The proof that in the calorimetric case at hand the simple standard BW shape
we have used is perfectly well suited is lengthy. We relegate it to  
Appendix \ref{AppTails}.

\section{The shortcomings of theory}

The precise treatment of a process in an atom with 67 electrons is obviously intricate. 
One limitation, in the case of EC, is the use of the sudden approximation.
While it is justified in the analysis of electrons captured from the inner orbitals, 
it is less so for captures from the $n\geq 3$ levels of interest here. 

In the sudden approximation (e.g.~in a $\beta$-decay) and in estimating secondary-hole probabilities, 
 the time required for the change of nuclear charge is assumed to be negligible relative to the typical atomic orbital times: the time it takes them to ``react'' to the new environment.
 In EC this assumption --used for decades without hardly a comment--
  is extended to the comparison
 of the time required to propagate
  the information that the captured electron has disappeared, relative to
 the rest of the characteristic atomic times: the former must be negligible.

The opposite to the sudden extreme is the adiabatic case, in which the electronic orbitals have time aplenty to slowly evolve from the parent-atom eigenstates to
the daughter-atom ones. In the extreme adiabatic limit the probability of
making ``second holes'' would vanish.

\subsection{The nuclear capture time}

The characteristic time, $\tau_{N}$, for the EC
 process $p+e\to n+\nu$ underlying the nuclear transition
is the inverse of the energy transfer $q_0$. For a process with ${\rm Q}\ll m_e$, such as the
one under discussion, $q_0\simeq m_e$ and $\tau_{N}\simeq 1/q_0$ is orders of magnitude
smaller than any characteristic atomic time. In the nuclear sense, EC is instantaneous.

\subsection{Our two-electron approach}

In the simple approach that we have discussed in Section \ref{theory}, two electrons (the captured one
and the one that is potentially exiting its orbital) play a singular role. The rest of the
electrons are only there to imply some effective values of $Z$ or to forbid some transitions.
But once more, a non-relativistic
Coulombic approximation\footnote{We used the limit $v/c\to 0$ everywhere but
in allowing for the particularly large effect of EC from the $l=1,\,j=1/2$
orbitals.}
provides guidance, which
in the case of the adequacy of the sudden approximation turns out to be quite relevant.

The relativistic retardation of signals implies that the sudden nuclear
transition $Z\!\to\! Z-1$ cannot be instantaneously felt by an electron
in an orbital of mean radius $r$. It takes a time of order $r$ for the orbital
to be ``informed''. Similarly, the absence of the electron that was captured is felt by
another atomic electron after a time of the order of the larger of the two mean
orbital radii.

Recall that the mean radius and binding energy of a Coulombic eigenfunction, in the usual 
notation, are:
\begin{eqnarray}
r[n,l]&=&{[3\,n^2-l(l+1)]/({2 \,m_e}\,Z\,\alpha)}\label{radius}\\
E[n]&=&{m_e\,(Z\,\alpha)^2/ (2\,n^2)}\label{energy}
\end{eqnarray}

The quantum-mechanical characteristic time for a bound state to ``react'' to a perturbation is 
the inverse of its binding energy. Thus,
deviations from the sudden approximation in a transition amplitude are, up to a factor of ${\cal O}(1)$, governed by a figure of merit $\delta$,
the ratio of the ``retardation" time to the quantum-mechanical ``reaction" time:
\begin{equation}
\delta\sim E\times r ={\cal O}( Z\,\alpha)
\label{delta}
\end{equation}
For $\delta\ll 1$ the sudden approximation is justified, while $\delta\gg 1$ corresponds to
the adiabatic limit.

In numerical examples, let us use the radius of an $l=0$ orbital as a characteristic
light-travel time, $t=r[n,0]$. The ``retardation time'' for another S-wave orbital, $n'$, to
be informed that the capture process has taken place is of order $t_r=\max(t[n],t[n'])$,
so that $\delta=E[n']\,t_r$.

For an M1 capture and a putative second hole in N1, $\delta=3 Z\alpha/4$,
that is a worrisome $\delta=0.24$ for $Z_{\rm eff}=43.2$. But
for the energies and mean radii obtained in a non-relativistic Hartree-Fock calculation,
$E(\rm N1,Ho)=0.414$ keV, $\langle r(\rm N1,Ho)\rangle=0.29$ \AA, and $E\times r\sim 0.062$,
a much smaller result. After a brief scare, the sudden approximation appears once again to be justified.

In our two-active-electron approach the use of (suddenly) overlapping wave-function is safe,
but the wave functions themselves are crude approximations. Thus the potential interest
of more ambitious calculations.

\subsection{A many electron system}

The methods of Faessler and collaborators \cite{Faessler1,Faessler2} are 
relativistic\footnote{The 
corrections to a non-relativistic approach are of ${\cal O}(Z\alpha)$,
like the effects neglected by use of the sudden approximation. 
}, 
and consistently use
Slater determinants to describe the fully anti-symmetrized states of the Ho
and Dy* atoms. In each atom the individual orbitals span an orthonormal set.
For Dy* this implies that {\it all} of its electrons must have had ample time,
after the nuclear capture, to readjust to the new situation.

Consider the capture of an M1 electron; the mean radius of
its orbital  is $\langle r(\rm M1,\rm Ho)\rangle=0.13$ \AA, which we will take as the pertinent
time for the capture information to arrive to three inner orbitals. For the other
electron in the M1 state, whose binding energy is
$E(\rm M1,\rm Ho)=2.05$ keV, the figure of merit is $\delta=E({\rm M1})\, r({\rm M1})\sim 0.135$.
For an L1 electron,  
$E(\rm L1,\rm Ho)=9.05$ keV and $\delta=E({\rm L1})\, r({\rm M1})\sim 0.6$.
Finally, for a K electron
$E(\rm K,\rm Ho)=55.6$ keV, and $\delta =E({\rm K})\, r(\rm M1)\sim 3.7>1$.

We conclude that the sudden approximation is not good for the Dy* atom 
in its ensemble. This statement is very specific to cases in which the captured
electron belongs to a rather external orbit: M1 or higher in the case of $^{163}$Ho.
Should we have discussed K-capture, the characteristic radius (or time) to spread
the information would have been a K radius and the energies of the orbits in which
second holes may be produced a K binding energy, or a smaller one. The sudden-approximation
figure of merit, $\delta$, would have always been significantly smaller than unity, justifying the sudden approximation.

\section{Platinum}

We have only discussed $^{163}$Ho. But measurements of
the next-to-best electron-capturing isotope 
\cite{ADR}, 
$^{193}$Pt, could be of help in understanding the difficult theoretical
issues associated with the
two-hole contributions. The problem is simply one of QED, like in the case of holmium
or the slightly harder problem of understanding  the human brain. 

\section{Conclusions}

The current very preliminary data on the calorimetric spectrum of $^{163}$Ho decay,
 in the region of the N peaks,
appear to be qualitatively described by the simplified theory we have discussed.
A similarly preliminary conclusion is that the description of the observations 
becomes quantitively satisfactory when the various theoretical single- and double-hole features 
are renormalized to fit the data. Once the data improve, we expect this conclusion to
become more solid.

On the way to measure or limit the neutrino mass, the relevant energy domain is 
not yet explored. It is the
one extending from the M2 and M1 single peaks towards the spectral endpoint,
the interval we have analized in detail. It is there that future data 
and the comparison of its features with theory will be of utmost interest.

We argued that,  relative to the ``old theory" with only
single-hole contributions,
double-hole effects may enhance the spectral
endpoint region by a factor of $\sim \! 40$. 
If this is correct, the statistical sensitivity on $m_{\nu}$ --which varies as the fourth 
root of the number of decays--
would improve by a factor $\sim \! 2.5$. 
We showed that the pileup problem would also be reduced by 
an even larger factor.

We cannot indubitably trust the above optimistic conclusions. Thus, we have
also studied the possibility that double- and single-hole contributions be
of similar magnitude close to the endpoint. This choice makes
the theoretical analysis of the data more challenging. 
But we have argued that also in this
very pessimistic case the introduction of just one extra parameter --besides
$m_\nu$, Q and the slope of a linear function of $E_c$-- should be enough
to analyse the data along well-trodden paths.

An issue that remains to be investigated is the possible existence, in a given
calorimeter's substrate, of BEFS oscillations. These would be due 
to Ho-decay electrons undergoing reflections in the crystal lattice \cite{BEFS}.

The possibility of a significant contribution of electron shake-ups,
leading to double-hole spectral peaks, was originally presented as
very damning \cite{Robertson1,Robertson2}. Somewhat ironically, the 
likely existence of a significant contribution of electron shake-offs
may be a very welcome blessing.

\subsection*{Acknowledgments}
We are indebted to Loredana Gastaldo, Angelo Nucciotti and Michael Rabin for fruitful discussions.
We thank the anonymous referee for very valuable comments and suggestions.
ADR acknowledges partial support from the  European Union FP7  ITN INVISIBLES (Marie Curie Actions, PITN- GA-2011- 289442).

\newpage

\appendix

\section{Interferences}
\label{app:interferences}

In the single-hole spectrum of Eq.\,(\ref{dRdEc}) quantum interferences have been omitted.
To discuss how good an approximation this is, we concentrate on the potentially most significant
interference: the one between the M1 and N1 amplitudes. For them to interfere, the $j=\rm M1,N1$
holes must decay into the same (generally two-hole) final state $f$. Introduce the notation by
writing the corresponding amplitude
and probability as:
\begin{eqnarray}
A_j^f &\propto& \phi_j(0) \; \frac{\sqrt{\gamma_j / \pi}}{ E-E_j + i\;\gamma_j}  
\sqrt{b(j \to f)}\;e^{{i\psi_j^f}}\;, \label{int1} \\
\Gamma^f &\propto& |\sum_j A_j ^f|^2 = 
\sum_j |A_j^f|^2\;+\; \sum_{k\neq j} A_k^f\,(A_j^f)^*\;,
\label{int2}
\end{eqnarray}
where $\gamma_j\equiv\Gamma_j/2$ is the half-width, $b(j\! \to\! f)$ is the branching fraction for this decay and $\psi_j^f$ is the unknown phase. 
Summing over all possible $f$ in Eq.\,(\ref{int2}), the first (second) term is the single hole expression 
(the interference). The explicit expression for the latter is:
 \begin{widetext}
\begin{equation}
2\, \phi_{M}(0) \phi_{N}(0) {\sqrt{\gamma_{M} \gamma_{N}}\over\pi}\;
\beta\,
\frac{[(E-E_{M})(E-E_{N})+\gamma_{M} \gamma_{N}] \cos(\psi)+ [\gamma_{M}(E-E_{N})-\gamma_{N}(E-E_{M})] \sin(\psi)} 
{[(E-E_{M})^2+\gamma_{M}^2]\;[(E-E_{N})^2+\gamma_{N}^2]} \, ,
\label{inter}
\end{equation}
 \end{widetext}
where we have suppressed the $c$ in $E_c$, the
1 in M1 and N1, $\beta$ stands for $\sum_{f} \sqrt{b(M \to f)\;b(N \to f)}$ and $\psi$ is the residual unknown phase. 

\begin{figure}[htbp]
\begin{center}
\includegraphics[width=0.48\textwidth]{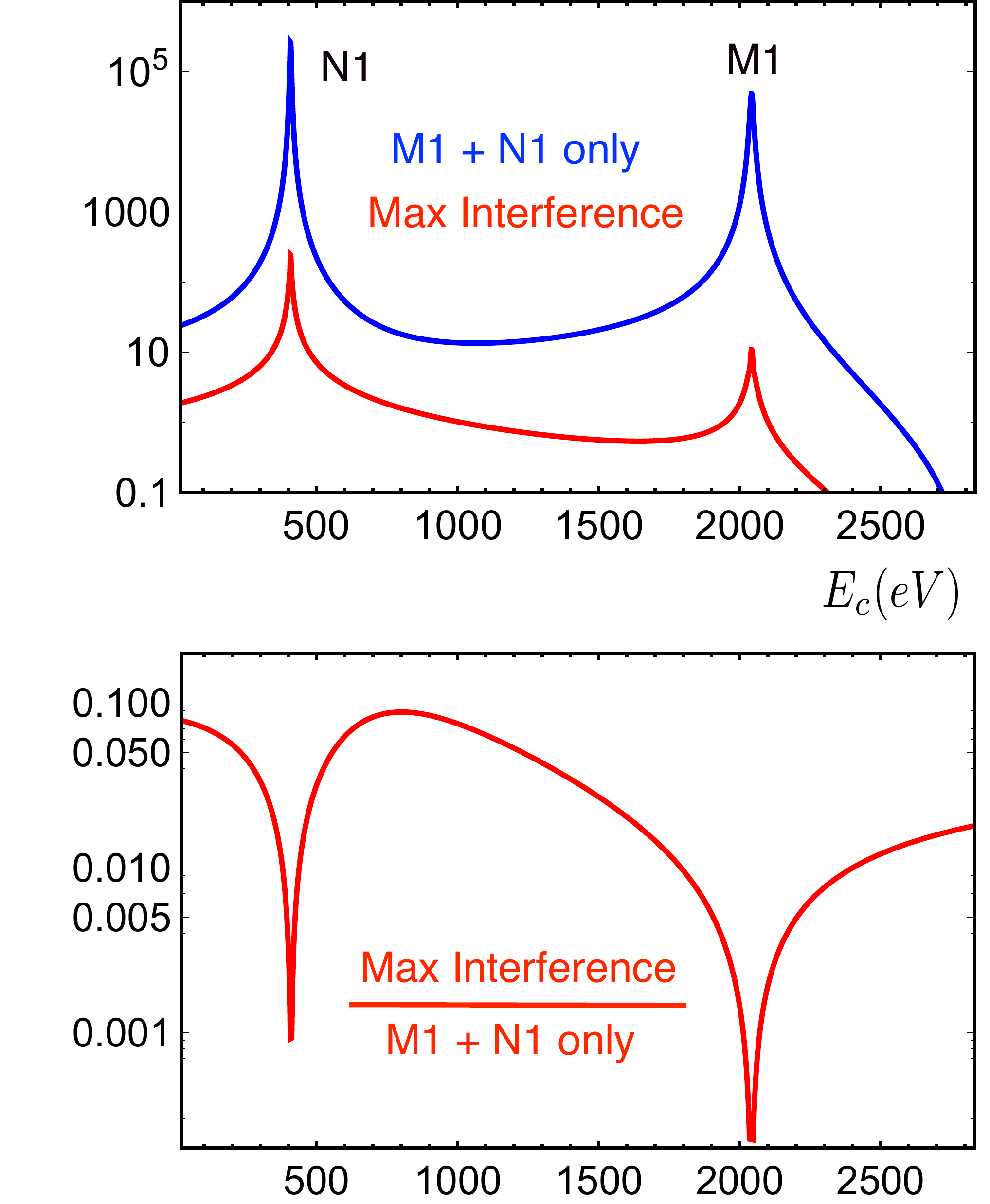}
\end{center}
\caption{Generous upper limits to the absolute value of the M1-N1 interferences,
for a hypothetical spectrum with only single M1 and N1 captures.
{\bf Above}:
The spectrum and the maximum interference.
{\bf Below}: The ratio of the maximum interference term to the non-interfering
spectrum.
}
\label{fig:Interferences}
\end{figure}

In searching for an upper limit to the interference term, we substitute $\beta$ by the larger
number $[{\sum_{f} b(M \to f) \times \sum_{f} b(N \to f)}]^{1/2}\sim 0.088$,  where the numerical
value is from \cite{mac}. Moreover, in Eq.\,(\ref{inter}) we take either $\cos(\psi)$ or $\sin(\psi)$
to be unity, depending on which of the two, at a given $E$, has the coefficient with the
largest absolute value.
The results of Eqs.\,(\ref{int2},\ref{inter})
are used to draw Fig.\,(\ref{fig:Interferences}), in whose lower panel we
see that $\sim 2$\% would be a generous upper limit for the fractional contribution
of the interference term close to the endpoint. 

The conclusion is that interferences can be neglected in the endpoint analysis, more so
if two-hole effects are dominant there.

\section{The tails of resonances}
\label{AppTails}

In writing, in the usual fashion, the contribution to the calorimetric spectrum
of a given hole of (positive) binding energy 
$E_H$, we have 
first employed the negligible-width approximation and the 
two-body phase space, $dN_2$, for the process $\rm Ho\to Dy^*+\nu$ to write:
\begin{equation}
dN_2/dE_\nu \propto p_\nu\, \delta(E_\nu-Q+E_H) \,E_\nu,
\end{equation}
where the square of the capture matrix element is $\propto E_\nu$.
Since in this case $E_c=E_H$ one simply has:
\begin{equation}
{dN_2\over dE_c} \propto \sqrt{(Q-E_c)^2-m_\nu^2}\, \delta(E_c-E_H) \,(Q-E_c)
\label{twobodyPS}
\end{equation}
The contribution of a hole of non-zero width $\Gamma_H$ can then be obtained by substituting
the $\delta$ function by a BW  to get, for the sum of single-hole contributions,
 Eqs.\,(\ref{dRdEc}-\ref{Eandp}). 

The purpose of the above naive reminder is to discuss the extent to which Eq.~(\ref{BW}) is
a good approximation, specially very many widths above a given resonance, in particular
the one corresponding to an M1 hole, whose contribution dominates the end of the spectrum.
In this respect, two related items need to be discussed, both concerning the single-hole contributions
to the calorimetric spectrum. The first is whether or not the two-body phase space is the
correct one to use, in spite of the fact that Dy* may decay to its ground state
by photon emission, in which case the process appears to have one extra 
final-state body (the photon). More often, the Dy* de-excitation starts with a complicated
chain of electron emissions, an apparently many-body final state. In all cases the
question arises: is the two-body Breit-Wigner shape of Eqs.~(\ref{dRdEc},\ref{BW}) adequate far away from its peak?

Consider, as a ``sanity test'', the possible but relatively very improbable case in which
the hole made by EC in the neutral daughter atom, Dy*, decays to the Dy
ground state by having the Dy* outermost electron transit to the hole, with the emission
of a photon. This is a process in which $E_c=E_\gamma$, and its phase space, $dN_3$, is a
three-body one: $\rm Ho \to Dy+\nu+\gamma$. Relative to Eq.~(\ref{twobodyPS}),
the calorimetric spectrum 
 acquires two extra factors:
\begin{equation}
{dN_3/ dE_c} \propto  E_c \, |M_\gamma|^2\, {dN_2/ dE_c},
\label{gamma}
\end{equation}
where  $E_c=E_\gamma$ arises from the photon's phase-space and
$M_\gamma$ is the photon emission matrix element.

The $E_c$ dependence in Eqs.~(\ref{BW}) differs from that in Eq.~(\ref{gamma}).
This is a well known situation \cite{PDG}, generally ``remedied'' by ``incorporating'' the kinematics (the extra factor of $E_c$) or the kinematics and dynamics
($E_c\,|M_\gamma|^2$) into an effective width.

Suppose we adopt the first of the two mentioned remedies by substituting
\begin{equation}
\Gamma_{\rm H}\to \widetilde\Gamma_{\rm H}=\Gamma_{\rm H}\, E_c/E_{\rm H}.
\end{equation}
Take the example of M1 capture ($E_{\rm M1}\simeq 2048$ eV).
The modification of the width in the denominator of the BW expression is 
immaterial at the endpoint, for $(Q-E_{\rm M1})^2\gg \widetilde \Gamma_{\rm M1}^2/4 $.
But the modification of the width in the numerator would alter the naive result
from Eq.~(\ref{BW}) by a factor changing linearly from 1 at $E_c=E_{\rm M1}$ to 
$Q/E_{\rm M1}\simeq 1.37$ at $E_c=Q$. 
This seems to be worrisome.

\begin{figure}[htbp]
\begin{center}
\includegraphics[angle=90,width=0.45\textwidth]{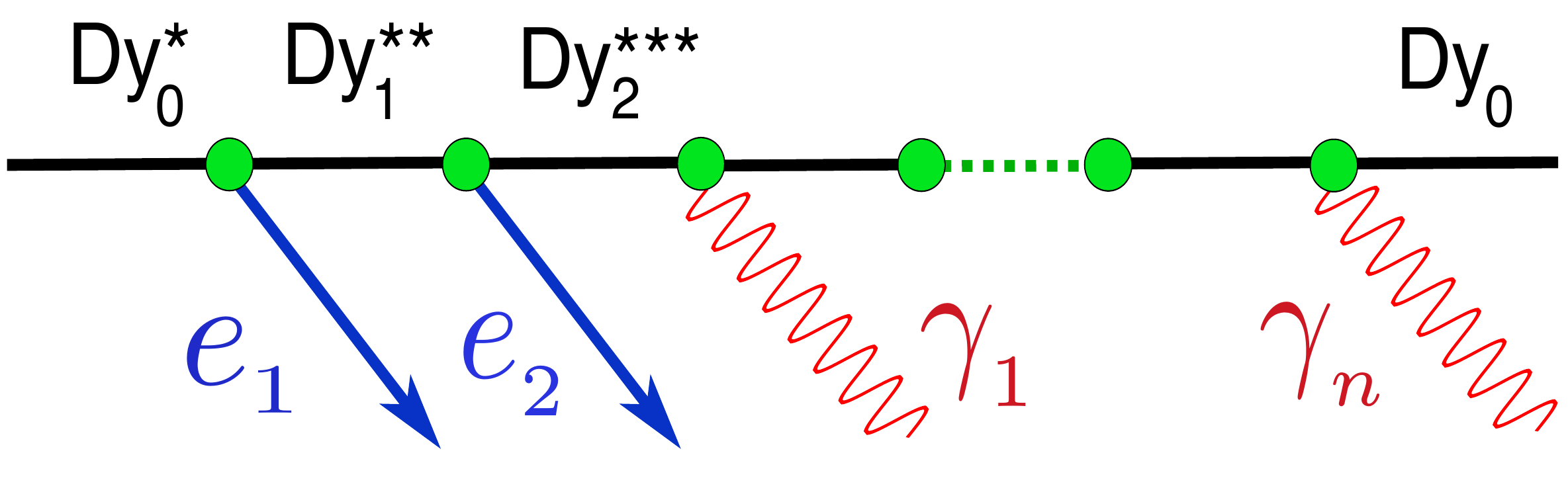}
\end{center}
\caption{A dominant chain of decays in the process $\rm Ho \to \nu+Dy^*$; 
$\rm Dy^* \to\,e's\,\&\,\gamma's + Dy_0$, with $\rm Dy_0$ the ground state 
of dysprosium. The subscripts of Dy are electric charges and the stars the
number of holes.}
\label{fig:Auger}
\end{figure}

As it turns out, in the  specific case at hand, the traditional expression naively based
on two-body phase space considerations, Eqs.~(\ref{dRdEc},\ref{BW}), is the correct one to use.
One reason is that the single-photon Dy* decay we discussed --which would require
a three-body phase space treatment-- has a negligible branching ratio. The
dominant decays consist of a series of Auger or Coster-Kronig transitions 
(in which the neutral Dy* emits electrons, becoming an increasingly charged ion),
followed by 
radiative 
transitions in which an outer atomic electron --or one coming from the  
negatively charged atomic neighborhood-- drops into the Dy-ion holes
and a photon is emitted. 
These processes are depicted in Fig.(\ref{fig:Auger}).

Consider any of the electron emissions in Fig.(\ref{fig:Auger}). The corresponding 
phase space expression contains a factor $p_e$, the momentum of the outgoing electron.
But, at the relevant very non-relativistic energies, this factor, to a very good approximation,
is compensated by the ``Fermi-function'', $F\sim 1/p_e$, which reflects the fact that the 
wave function of the outgoing electron is not that of a free particle,
but the one of an electron subject to the field of a charged ion. For any of the soft photon emissions
in Fig.(\ref{fig:Auger}) the pertinent width may be modified by a remedial factor 
$E_\gamma/\langle E_\gamma\rangle$, but that function differs very little from unity in its
narrow allowed range (unless $E_\gamma^{\rm max}\sim Q$, the unlikely case discussed in 
the next paragraph).

The emission of only one photon is highly suppressed, as we said above. There still remains the possibility that one of the (many) photons emitted in the decay --as depicted in Fig.(\ref{fig:Auger})--
carries all of the energy $E_c\simeq Q$, requiring a modified width in the corresponding Breit-Wigner. 
 But in that case all the other emitted photons have  momenta $p\simeq 0$
and the phase space (multiply) vanishes. The multi-body phase space has a multidimensional
pole at the central energies of each of the transitions. At $E_c\simeq Q$ the overwhelmingly
most likely situation corresponds to all energies being close to this multiple pole and adding up
to $E_c\simeq Q$. All in all the overall process is described by the naive two-body phase space
expression of Eq.~(\ref{BW}), with its width unmodified.

To conclude: an event in the process we are discussing can 
be viewed as the two-body decay of the calorimeter ``before'' to a neutrino and the 
unstable ``excited calorimeter''
immediately after. The calorimeter then releases its excess energy
--which is calorimetrically recorded--
to return to a new ground state (with one fewer Ho and one extra Dy atoms).

\end{document}